\documentclass[lettersize,journal]{IEEEtran}
\usepackage{amsmath,amsfonts}
\usepackage{algorithmic}
\usepackage{array}
\usepackage[caption=false,font=normalsize,labelfont=sf,textfont=sf]{subfig}
\usepackage{textcomp}
\usepackage{stfloats}
\usepackage{url}
\usepackage{verbatim}
\usepackage{graphicx}
\def\BibTeX{{\rm B\kern-.05em{\sc i\kern-.025em b}\kern-.08em
    T\kern-.1667em\lower.7ex\hbox{E}\kern-.125emX}}
\usepackage{balance}
\usepackage[table]{xcolor}
\usepackage[numbers]{natbib}
\definecolor{tinygray}{gray}{0.94}
\definecolor{lightgray}{gray}{0.90}

\usepackage{amsmath}
\usepackage{hyperref}

\begin{document}
\title{CoEditor++: Instruction-based Visual Editing via Cognitive Reasoning}
\author{Minheng Ni, Yutao Fan, Zhengyuan Yang, Yeli Shen, Yuxiang Wei, Yaowen Zhang,\\Lijuan Wang, \textit{Senior Member, IEEE},  Lei Zhang, \textit{Fellow, IEEE}, and Wangmeng Zuo, \textit{Senior Member, IEEE}
\thanks{Minheng Ni, Lei Zhang are with the Hong Kong Polytechnic University, Kowloon 999077, Hong Kong SAR (e-mail: minheng.ni@connect.polyu.hk, cslzhang@comp.polyu.edu.hk).

Minheng Ni, Yutao Fan, Yeli Shen, Yuxiang Wei, Yaowen Zhang, Wangmeng Zuo are with the Harbin Institute of Technology, Harbin 150001, China (e-mail: mhni@stu.hit.edu.cn, fanyutao@stu.hit.edu.cn, ylshen@stu.hit.edu.cn, yuxiang.wei.cs@gmail.com, ywzhang@stu.hit.edu.cn, csxmli@gmail.com, wmzuo@hit.edu.cn).

Zhengyuan Yang, Lijuan Wang are with the Microsoft, Redmond 98052, United States (e-mail: zhengyang@microsoft.com, lijuanw@microsoft.com).}}

\markboth{Journal of \LaTeX\ Class Files,~Vol.~18, No.~9, September~2020}%
{CoEditor++: Instruction-based Visual Editing via Cognitive Reasoning}

\maketitle

\begin{abstract}
Recent advances in large multimodal models (LMMs) have enabled instruction-based image editing, allowing users to modify visual content via natural language descriptions. However, existing approaches often struggle with high-level semantic reasoning and visual consistency, particularly under ambiguous or complex instructions. To address these challenges, we propose CoEditor++, a cognitively structured, training-free framework that decomposes editing into ``\textit{what to edit}'' and ``\textit{how to edit}'' through two cognitive stages with a reflective self-selection mechanism, enabling robust, fine-grained, and interpretable editing. Built entirely from open-sourced components, CoEditor++ requires no additional training or fine-tuning, ensuring transparency and cross-domain applicability. We evaluate CoEditor++ on SmartEdit, a widely used benchmark for general editing, and AltBear, a privacy and compliance-oriented benchmark. Experimental results show that CoEditor++ achieves state-of-the-art performance in both general editing and responsible editing tasks compared with open-sourced models that require training on specialized editing datasets maintaining significantly higher visual consistency. When compared with closed-source models such as Nano Banana Pro or GPT-4o, CoEditor++ preserves comparable instruction following while still substantially outperforming them in visual consistency. Extensive ablation studies confirm that the effectiveness of CoEditor++ benefits from its structured cognitive design rather than any specific model component. Our findings suggest the potential toward cognitive-centric instruction-based image editing. We release our code and model at \hyperlink{https://xxx.com}{https://xxx.com}.

\end{abstract}


\section{Introduction}

In recent years, large multimodal models (LMMs) have made remarkable progress, substantially enhancing machines' abilities to perceive~\citep{cai2025pixel, wang2025pixelthink, su2025pixel}, reason about~\citep{team2025kimi, chris2504skywork, wang2025think}, and generate complex visual content~\citep{wang2025ovis, chen2025unicode, ai2025ming} under natural language instructions.
Among the emerging capabilities, modifying images based on user-provided textual descriptions \textit{i.e.}, instruction-based image editing~\citep{fang2025got, wei2024omniedit, liu2025step1x}, has attracted growing attention for its wide-ranging applications, including creative design~\citep{venkatesh2025crea}, content moderation~\citep{wang2024moderator}, accessibility~\citep{chang2024editscribe}, compliance~\citep{meding2025constitutes}, fairness correction~\citep{ni2024responsible}, and privacy protection~\citep{wang2025edit}. 
This paradigm allows users to express abstract or high-level intent via natural language, bridging the gap between human imagination and visual manipulation.

\begin{figure}[t]
\centering
\includegraphics[width=\linewidth]{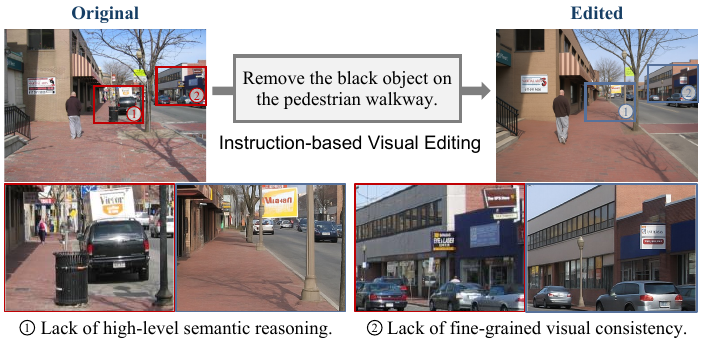}
\caption{\textbf{Example of high-level semantic reasoning and fine-grained visual consistency.} Although the model followed the instructions, it failed to accurately identify the content that needed to be modified and made unnecessary changes to the image background. These problems will become increasingly pronounced during continuous editing.}
\label{fig:intro}
\end{figure}

Despite recent progress, current methods~\citep{xu2025insightedit, zhang2025context, ma2025x2i} still struggle with several persistent challenges, especially in scenarios which need high-level semantic reasoning, or fine-grained visual consistency as shown in Figure \ref{fig:intro}. 
%
%
%
When an edit fails, it is difficult to determine whether the cause lies in a failure of understanding or generation control.
First, most models lack a structured reasoning mechanism to decompose abstract instructions into actionable editing plans. 
As a result, they often fail to generalize across varying abstraction levels or diverse visual contexts. 
Second, existing methods typically process the entire image holistically, without explicit region isolation mechanisms.
This often causes unnecessary changes to irrelevant areas, particularly harmful in visually sensitive images such as those containing text or precise visual layouts. 
These problems will become increasingly pronounced during continuous editing, eventually making results unacceptable.

To solve these problems, we propose CoEditor++, which draws inspiration from dual process theory in cognitive science rather than attempting to train a stronger end-to-end model. CoEditor++ treats image editing as a complex task that requires the involvement of System 2, logical and deliberative reasoning, rather than merely System 1, intuitive and straight-forward pixel transformation.
Built entirely from open-source components, an instruction-based segmentation, inpainting, and large-multimodal model, CoEditor++ embodies our central hypothesis: structured cognitive coordination can unlock powerful editing capabilities even from general-purpose models. 
Unlike methods reliant on task-specific datasets or costly fine-tuning, our approach emphasizes interpretability, modularity, and cross-task adaptability.

Specifically, CoEditor++ models image editing as two cognitive stages that incorporate perception, planning, and reflection. 
For stage I, localization cognitive process (LCP) identifies the image regions relevant to the instruction.
Unlike static heuristics, LCP performs dynamic multimodal cognitive analysis, simulating human-like attentional mechanisms to locate target areas.
This helps to precisely locate the target editing region, even when spatial cues are absent.
For stage II, modification cognitive process (MCP) determines how to modify the identified regions while preserving global semantics and visual realism. 
Also rooted in multimodal reasoning, MCP flexibly infers description of target content via LMM and applies appropriate edits via inpainting model under both explicit and underspecified instructions. 
Additionally, we introduce a reflective self-selection mechanism in both stages. 
During each reasoning round, the LMM is adopted to select the best candidate via unsupervised evaluation. 
Such a reflective loop improves robustness under ambiguous instructions and reduces error accumulation across editing steps. 

We evaluate CoEditor++ on two comprehensive benchmarks: SmartEdit~\citep{huang2024smartedit}, a widely used suite covering general editing tasks that require complex reasoning and spatial grounding, and AltBear~\citep{ni2024responsible}, which we previously introduced to focus on privacy and ethical compliance.
Together, these benchmarks form a rigorous testing suite that balances technical precision with societal relevance.
In general editing, CoEditor++ introduces virtually no pixel-level changes to unedited regions, significantly outperforming mainstream models including GPT-4o~\citep{hurst2024gpt}, Nano Banana~\citep{Nano_Banana}, and Nano Banana Pro~\citep{Nano_Banana_Pro}. 
In terms of editing success rate, CoEditor++ performs on par with closed-source models, \textit{e.g.}, GPT-4o or Nano Banana Pro, and surpasses all other open-source models with specialized training data such as ICEdit~\citep{zhang2025context}, Seed-X~\citep{ge2024seed}, and Qwen-Image-Edit~\citep{wu2025qwen}. 
In addition, CoEditor++ demonstrates strong generalization ability with a clear advantage on the AltBear benchmark. 
Qualitative results show that CoEditor++ aligns consistently with user intent, maintains high visual realism, and avoids unnecessary collateral modifications. 
Most importantly, CoEditor++ preserves consistency across multi-round editing and prevents over-modification, making it suitable for high-fidelity real-world deployment.

Ablation studies further demonstrate that CoEditor++‘s effectiveness stems not from any single component, but from its structured reasoning design. For instance, even with ground-truth masks and powerful inpainting models, success rates remain extremely low; naive fusion of segmentation and inpainting without reasoning fails almost entirely. This reinforces our insight: instruction-based editing is a reasoning-centric problem, not merely a combination of segmentation and generation. Moreover, replacing our open-source LMM with closed-source models, \textit{e.g.}, GPT-4o or Nano Banana Pro, produces similar results, confirming that our performance stems from the framework’s structure, not any specific model.

This work substantially extends our ECCV 2024 paper, Responsible Visual Editing~\citep{ni2024responsible}. 
While the original system focused on responsible editing, this study generalizes the task to a broader instruction-based setting, introducing a cognitively structured, model-agnostic framework. 
Beyond conceptual expansion, we fully replace closed-source components with open-source modules, enabling transparent and reproducible deployment. 
We also introduce a new benchmark, conduct deeper experiments, and present extensive ablations and visualizations. 
In summary, our key contributions are:
\begin{itemize}
\item We propose CoEditor++, a cognitively structured framework that decomposes editing into ``\textit{what to edit}” (LCP) and ``\textit{how to edit}” (MCP), enhancing interpretability and modular generalization with a reflective self-selection mechanism at both stages to improve robustness;
\item CoEditor++ is built entirely from open-source components, requiring no training or fine-tuning, ensuring reproducibility and cross-domain applicability;
\item We evaluate CoEditor++ on SmartEdit and AltBear benchmark across general and responsible editing tasks, showing CoEditor++ outperforms closed-source models, \textit{e.g.}, GPT-4o or Nano Banana Pro in visual consistency while matching it in task success rate;
\item Extensive ablation and visualization studies confirm that CoEditor++’s performance arises from its cognitive reasoning structure, not from any individual module, with human-like editing paths supporting its interpretability.
\end{itemize}

Our findings highlight the potential for instruction-based editing to move beyond data-centric model training, toward cognitively guided, modular architectures that combine flexibility, transparency, and practical utility.

\section{Related Work}

\subsection{Instruction-based Image Editing}

In recent years, with the advancement of generative models~\citep{kingma2013auto, goodfellow2014generative, croitoru2023diffusion},  instruction-based image editing has made rapid progress. Pioneering works such as InstructPix2Pix~\citep{Brooks_2023_CVPR} formally introduced the task of editing real images using natural language instructions, achieving this goal by training conditional diffusion models~\citep{ho2022classifier} on large-scale synthetic datasets. As the limitations of synthetic data quality have become increasingly apparent, subsequent research has gradually shifted towards enhancing the realism of the data. For example, MagicBrush~\citep{zhang2023magicbrush} significantly improved model performance and instruction adherence via annotated data training. Models such as HQ-Edit~\citep{hui2025hq}, UltraEdit~\citep{zhao2024ultraedit}, and AnyEdit~\citep{yu2025anyedit} have further advanced scalable data generation processes, proposing large-scale, high-quality training pairs and emphasizing the decisive role of high-quality data in enhancing editing capabilities. Recent studies have explored the integration of large multimodal models (LMMs) into end-to-end architectures. For instance, MGIE~\citep{fu2023guiding} and SmartEdit~\citep{huang2024smartedit} leverage multimodal large models to understand instructions and achieve fusion of textual and visual features, leading to unified models like OmniGen~\citep{xiao2025omnigen} and Q wen-Image-Edit~\citep{wu2025qwen}. These models can handle diverse generative tasks and reflective output optimization using a single Transformer. Although existing instruction-based image editing models exhibit powerful capabilities after extensive data training, they struggle to generalize to complex scenarios and find it challenging to ensure image consistency due to the lack of explicit cognitive processes.

\subsection{Large Multimodal Models}

By integrating visual understanding modules into LLMs, large multimodal models have gradually demonstrated strong multimodal reasoning and cognitive abilities. As one of the pioneering works in LMMs, LLaVA~\citep{liu2023visual} introduced visual instruction tuning, a paradigm that greatly promoted vision-language alignment. Subsequent models such as the Qwen-VL series~\citep{bai2025qwen2, wang2024qwen2}, InternVL~\citep{chen2024expanding, zhu2025internvl3}, and Ming-Omni~\citep{ai2025ming} further extended these capabilities, excelling in precise object localization, long video understanding, and handling multiple input modalities including text, images, audio, and video. To advance from basic understanding to more complex multi-step reasoning, the inclusion of detailed chains-of-thought (CoT)~\citep{wei2022chain} in the reasoning process has driven further progress. Additionally, reinforcement learning (RL) has become a powerful paradigm for enhancing these reasoning abilities~\citep{guo2025deepseek, yu2025dapo, yue2025vapo}. Frameworks like VLM-R1~\citep{shen2025vlm} and Vision-R1~\citep{huang2025vision} systematically apply reinforcement learning with programmatically verifiable rewards to improve performance in general vision and complex mathematical reasoning tasks. Innovations in the field also include teaching models to perform selective reasoning to avoid unnecessary computation~\citep{wang2025think}, or employing external visual tools that enable direct reasoning in pixel space~\citep{wang2025pixelthink, su2025pixel, kang2025viki}. Despite the impressive capabilities of large multimodal models focused on reasoning, existing work has yet to achieve multimodal reasoning and cognitive abilities during the generative process, making it difficult to accomplish complex editing in zero-shot scenarios.
\begin{figure*}[t]
\centering
\includegraphics[width=\linewidth]{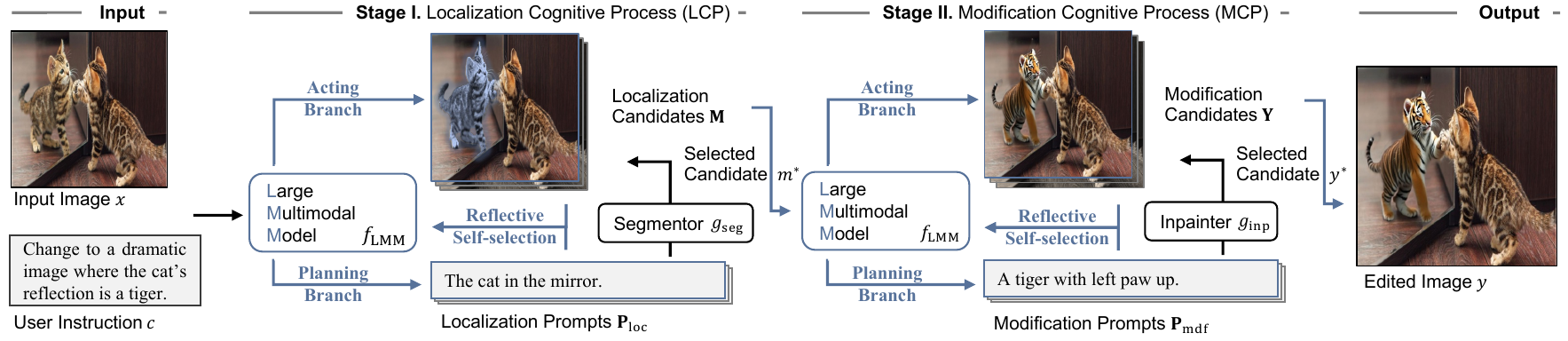}
\caption{\textbf{Overview of the CoEditor++ with two stages of cognitive process and a reflective self-selection to find out the best intermediate result.} In the localization cognitive process (LCP), a large multimodal model (LMM) jointly processes the input image $x$ and user instruction $c$ to generate a set of localization prompts $\mathbf{P}_{\mathrm{loc}}$, describing a candidate region each. These prompts are mapped to localization
candidates $\mathbf{M}$ via a segmentation model. In the modification cognitive process (MCP), the LMM formulates a set of modification prompts $\mathbf{P}_{\mathrm{mdf}}$ based on the selected mask $m^{*}$, which guides an inpainting model to synthesize modification candidates $\mathbf{Y}$ and select the final edited result $y^{*}$ via reflective self-selection. This framework explicitly decouples “\textit{what to edit}” and “\textit{how to edit}”, enabling robust, interpretable, and contextually aligned visual editing. }
\label{fig:overview}
\end{figure*} 

\section{Methodology}

In this section, we introduce the cognitively structured two-stage pipeline of CoEditor++, designed to address persistent challenges in instruction-based image editing. Inspired by the reflective and stepwise editing workflow of professional human editors, CoEditor++ systematically decomposes the editing process into two interacting stages: the localization cognitive process (LCP) and the modification cognitive process (MCP). Each stage incorporates planning, acting, and a reflective self-selection mechanism, enabling robust and interpretable reasoning in the face of ambiguous, high-level, or contextually diverse instructions. 

\subsection{Framework Overview}

Unlike traditional one-stage or monolithic approaches, CoEditor++ is designed as a modular pipeline that explicitly separates “\textit{what to edit}” from “\textit{how to edit}”. Specifically, LCP identifies the spatial regions in the image relevant to the instruction, while MCP determines the optimal content modifications within these regions. Both stages default to a single-candidate path, but can invoke a reflective self-selection mechanism to generate and evaluate multiple candidates, thereby simulating the iterative analysis and decision refinement process observed in expert human editing.

As illustrated in Fig.~\ref{fig:overview}, given an input image $x$ and a natural language instruction $c$, the framework first generates a localization prompt $p_{\mathrm{loc}}$ that describes the region semantically related to the instruction. This prompt is then mapped into a binary mask $m$ in the pixel space indicating the target editing region. Subsequently, the system generates a modification prompt $p_{\mathrm{mdf}}$ describing the specific editing requirement for the selected region, which then guides the content generation process to produce the edited image $y$. Meanwhile, the reflective self-selection chooses the best intermediate results.

\subsection{Localization Cognitive Process (LCP)}

The localization cognitive process (LCP) aims to answer the core question: \textit{Which part of the image should be edited according to the user’s instruction?} This stage is especially critical in scenarios where instructions are ambiguous, abstract, or lack explicit specification of the relevant region. Here, CoEditor++ cognitively translates user intent into spatially localized visual targets. Taking the image and textual instruction as input, it outputs a binary mask that highlights the regions relevant for editing. This mapping does not rely on keyword matching or static semantic segmentation, but instead performs dynamic multimodal reasoning, attending to visual and linguistic cues. Through this process, LCP ensures that subsequent modification operations are only applied to contextually appropriate regions, minimizing unintended changes to irrelevant areas, a property particularly important for images with complex structures and dense visual content. In essence, LCP anchors abstract language to spatial coordinates, providing a precise and robust basis for downstream editing even when instructions are abstract, ambiguous, or spatially vague.

Formally, given input image $x \in \mathbb{R}^{H \times W \times 3}$ and editing instruction $c$, LCP generates a localization mask $m \in \{0,1\}^{H \times W}$, where:
\begin{equation}
m_{i,j} = 
\begin{cases}
0 & \text{if pixel } (i,j) \text{ should remain unchanged}, \\
1 & \text{otherwise}.
\end{cases}
\end{equation}

\subsubsection{Planning Branch}

The planning branch simulates the initial semantic grounding performed by a human editor. Specifically, the system generates a natural language localization prompt $p_{\mathrm{loc}}$ to describe a subregion:
\begin{equation}
p_{\mathrm{loc}} = f_{\mathrm{LMM}}\left(x, c; \mathcal{S}_{\mathrm{loc}}\right),
\end{equation}
where $f_{\mathrm{LMM}}(\cdot)$ denotes a large multimodal model conditioned on the image $x$, instruction $c$, and a system message $\mathcal{S}_{\mathrm{loc}}$. The localization prompt $p_{\mathrm{loc}}$ explicitly bridges abstract linguistic expressions and spatial visual targets.

\subsubsection{Acting Branch}

After obtaining the localization prompt $p_{\mathrm{loc}}$, the acting branch maps it to an initial binary mask $\tilde{m}$ via a segmentation model $g_{\mathrm{seg}}$:
\begin{equation}
\tilde{m} = g_{\mathrm{seg}}\left(x, p_{\mathrm{loc}}\right).
\end{equation}
To further improve mask quality and reduce residual noise from segmentation, morphological dilation $\delta(\cdot)$ is applied:
\begin{equation}
m = \delta\left(\tilde{m}\right).
\end{equation}
This operation ensures that the resulting mask $m$ is more contextually informative and spatially coherent, mitigating issues such as mask fragmentation or omission of relevant content. Specifically, if the editing instruction involves adding new content (\textit{e.g.}, placing an orange on a plate), the LCP will select the affected region (\textit{e.g.}, the entire plate) as $m$.

\subsubsection{Reflective Self-selection}

Although the planning and acting branches work in tandem, localization errors (such as over-segmenting irrelevant regions or missing target areas) can still occur, especially under ambiguous or underspecified instructions. To address this, CoEditor++ introduces a reflective self-selection mechanism in the LCP stage, enabling the system to internally evaluate and refine localization results prior to editing.

This mechanism first prompts the LMM to generate diverse region descriptions $p_{\mathrm{loc}}$ for the same instruction-image pair $(x, c)$. The LMM then scores each description based on its alignment with the instruction and image, selecting the top-scoring $p_{\mathrm{loc}}^*$ to guide segmentation:
\begin{equation}
p_{\mathrm{loc}}^* = \arg\max_{p \in \mathbf{P}_{\mathrm{loc}}} f_{\mathrm{LMM}}\left(p|x, c; \mathcal{S}_{\mathrm{rfl}}\right),
\end{equation}
where $\mathbf{P}_{\mathrm{loc}}$ is the set of localization prompts.

Next, the selected description $p_{\mathrm{loc}}^*$ is used to generate multiple segmentation masks $m$, each of which is also scored by the LMM based on spatial accuracy and semantic relevance, with the best mask $m^*$ ultimately selected:
\begin{equation}
m^* = \arg\max_{m \in \mathbf{M}}f_{\mathrm{LMM}}\left(m|x, p_{loc}^{*}; \mathcal{S}_{\mathrm{rfl}}\right),
\end{equation}
where $\mathbf{M}$ is the set of all localization candidates.

This reflective process emulates how human editors compare and deliberate over different options before finalizing the editing region. Leveraging this reflective loop, CoEditor++ can achieve precise and instruction-aligned region selection, especially in visually complex or semantically subtle scenarios.

\subsection{Modification Cognitive Process (MCP)}

Once the target region $m^*$ has been localized, the MCP addresses the next core question: \textit{How should the selected region be modified to fulfill the instruction?} The MCP is responsible for generating new content or transforming existing elements within the selected region, ensuring that the result is both realistic and contextually coherent with the global image. The main challenge lies in translating high-level or underspecified instructions, \textit{e.g.}, make it look more elegant, into semantically faithful and visually harmonious modifications. The MCP is thus particularly important for handling abstract or incomplete instructions. Its structure mirrors that of the LCP, comprising a planning branch, which formulates a detailed and actionable editing plan based on the user instruction and localized region, and an acting branch, which invokes the inpainting model according to the editing plan and mask to realize the specific transformation, thereby ensuring that the output aligns with the user’s intent while maintaining consistency with the input image and avoiding unnecessary alterations.

\subsubsection{Planning Branch}

The planning branch generates a modification prompt $p_{\mathrm{mdf}}$, which provides a detailed and executable editing plan for the region $m^*$:
\begin{equation}
p_{\mathrm{mdf}} = f_{\mathrm{LMM}}\left(x, c, m^*; \mathcal{S}_{\mathrm{mdf}}\right),
\end{equation}
where, $p_{\mathrm{mdf}}$ may specify stylistic, structural, or content-level transformations, \textit{e.g.}, replace modern furniture with rustic wooden elements and apply a brownish tone. This intermediate representation guides the subsequent image generation, decoupling high-level reasoning from low-level synthesis.

\subsubsection{Acting Branch}

Given the modification prompt $p_{\mathrm{mdf}}$, the system synthesizes the edited image $y$ via a generative model $g_\mathrm{inp}$, conditioned on $x$, $m^*$, and $p_{\mathrm{mdf}}$:
\begin{equation}
y = g_\mathrm{inp}\left(x, m^*, p_{\mathrm{mdf}}\right).
\end{equation}
In practice, for addition-type instructions, the LCP can be re-executed after generation based on applying partial noise~\citep{ni2024ores} to obtain the most accurate $m^*$.

\subsubsection{Reflective Self-selection}

Similar with LCP, the planning branch generates diverse editing plans $p_{\mathrm{mdf}}$ for the image $x$, instruction $c$, and mask $m^*$. It then evaluates these candidates according to semantic alignment and visual context, selecting the optimal plan:
\begin{equation}
p_{\mathrm{mdf}}^* = \arg\max_{p \in \mathbf{P}_{\mathrm{mdf}}} f_{\mathrm{LMM}}\left(p \mid x, c, m^*; \mathcal{S}_{\mathrm{rfl}}\right),
\end{equation}
where $\mathbf{P}_{\mathrm{mdf}}$ is the set of modification prompts.

The acting branch produces multiple candidate edited images, \textit{e.g.}, by varying noise parameters and random seeds, and these candidates are again evaluated by the LMM in terms of semantic fidelity, visual quality, and consistency with the unedited regions, ultimately outputting the best result $y^*$:
\begin{equation}
y^* = \arg\max_{y \in \mathbf{Y}} f_{\mathrm{LMM}}\left(y \mid x, c; \mathcal{S}_{\mathrm{rfl}}\right),
\end{equation}
where $\mathbf{Y}$ is the set of all modification candidates.

\subsection{Implementation Details}

CoEditor++ requires no additional training and consists of open-source components only, ensuring reproducibility and adaptability. We employ Qwen2.5-VL-72B-Instruct as our multimodal reasoning backend. For localization, the language-guided segmentation model LISA-13B is used to generate binary masks, with a morphological dilation ratio of $20$ pixels. For modification, the diffusion-based inpainting model Flux-Inpainting is used, and the default number of reflective samples $N$ is set to $5$. We keep the original size of the image, and all experiments are conducted on a single NVIDIA H20.
\section{Experiment}

\subsection{Experimental Setup}

\paragraph{Datasets}
We conduct a comprehensive evaluation on two distinct benchmarks to assess CoEditor++'s capabilities across a wide spectrum of editing tasks. The first is SmartEdit~\citep{huang2024smartedit}, a widely adopted benchmark for general-purpose, instruction-based image editing. SmartEdit is designed to evaluate both spatial locating and semantic reasoning. It contains two tasks: Understanding and Reasoning. In the understanding task, images typically contain multiple objects, and the editing instructions require the model to accurately localize the correct target among distractors. This places high demands on spatial comprehension and multimodal alignment. In contrast, the reasoning task emphasizes abstract or underspecified instructions, where the editing goal is not directly stated but must be inferred from context. These tasks challenge the model’s ability to perform nuanced spatial understanding and higher-level reasoning. The second is AltBear, a specialized benchmark we previously introduced for evaluating performance on tasks demanding high responsibility and ethical consideration, such as unsafe content removal and fairness correction.

\paragraph{Evaluation}
For evaluation, both benchmarks provide inputs in the form of an original image and a text instruction. 
In SmartEdit, each sample is additionally accompanied by a ground-truth (GT) mask that precisely indicates the minimal region to be modified. These GT masks are used exclusively during evaluation to assess visual consistency by isolating the unedited areas. They are never exposed to any model.
In contrast, AltBear does not provide GT masks, since responsible editing tasks often lack a clearly defined visual target. For this reason, visual consistency on AltBear is assessed over the full image, but only on samples where GPT-4o deems the edit successful, ensuring that consistency metrics reflect collateral changes in high-quality outputs. We evaluate both the safety and fairness tasks within AltBear to cover different categories of responsible editing tasks. It is worth noting that images in the AltBear dataset trigger safeguards in closed-source models, which has also been mentioned in previous work~\citep{ni2024responsible}. As a result, we are unable to report their performance on this dataset.

\paragraph{Metrics}
Our evaluation protocol is designed to measure two critical dimensions of performance: visual consistency and instruction following. To quantify visual consistency,  the degree to which an edited image preserves the content of the original, we employ three standard metrics: Peak Signal-to-Noise Ratio (PSNR), Structural Similarity Index (SSIM), and Learned Perceptual Image Patch Similarity (LPIPS)~\citep{hore2010image, zhang2018unreasonable}. These are calculated against the unedited regions defined by the GT mask to specifically measure unwanted collateral modifications. Among these three metrics, except for LPIPS where lower is better, the other three metrics are better. To assess instruction following, how well the output aligns with the user's intent, we use the CLIP Score~\citep{radford2021learning} to measure the semantic similarity between the edited image and the text instruction. Critically, we also report the Success Rate (Succ), a metric determined by GPT-4o's automated evaluation of whether the edit successfully fulfills the instruction's objective.

\paragraph{Baselines}
We compare CoEditor++ against a comprehensive suite of state-of-the-art instruction-based image editors. This includes leading academic models such as InstructPix2Pix~\citep{Brooks_2023_CVPR}, MagicBrush~\citep{zhang2023magicbrush}, InstructDiffusion~\citep{geng2024instructdiffusion}, SmartEdit-13B~\citep{huang2024smartedit}, SEED-X~\citep{ge2024seed}, OmniGen~\citep{xiao2025omnigen}, Insight-Edit~\citep{xu2025insightedit},  ICEdit~\citep{zhang2025context}, and Qwen-Image-Edit~\citep{wu2025qwen}. To situate our performance against the most powerful closed-source proprietary methods, we also include GPT-4o~\citep{hurst2024gpt}, Nano Banana Pro~\citep{Nano_Banana_Pro}, and Nano Banana~\citep{Nano_Banana}, as a formidable reference. All models were run using their official codebases and recommended hyperparameters to ensure a fair comparison. Please note that Insight-Edit did not release their model or code, so we report all results from their original paper. Considering their framework and training data scale, we still categorize them as an open-source method.

\begin{table*}[!t]
\centering
\renewcommand\arraystretch{1.2}

\caption{\textbf{Comparison of PSNR / SSIM / LPIPS / CLIP / Succ across methods on the Understanding, Reasoning and Responsible tasks.} ``-" denotes no results reported in the original paper. Best results are in \textbf{bold}, second best are \underline{underlined}.}
\sc
\resizebox{\textwidth}{!}{
  \setlength{\tabcolsep}{3.65mm}
  \begin{tabular}{l|ccccc|ccccc|cccc}
    \hline
    \rowcolor{lightgray}
    \textbf{Methods}
      & \multicolumn{5}{c|}{\textbf{Understanding}}
      & \multicolumn{5}{c|}{\textbf{Reasoning}}
      & \multicolumn{4}{c}{\textbf{Responsible}}
    \\\hline\hline
      & \textbf{PSNR} & \textbf{SSIM} & \textbf{LPIPS} & \textbf{CLIP} & \textbf{Succ}
      & \textbf{PSNR} & \textbf{SSIM} & \textbf{LPIPS} & \textbf{CLIP} & \textbf{Succ}
      & \textbf{PSNR} & \textbf{SSIM} & \textbf{LPIPS} & \textbf{Succ}
    \\\hline
    \multicolumn{15}{c}{Closed-source Methods} \\
        \hline
        \rowcolor{tinygray}GPT-4o~\citep{hurst2024gpt}
      & 15.303 & 0.496 & 0.208 & 23.858 & 0.977
      & 15.125 & 0.435 & 0.217 & 20.657 & 0.867
      &    –   &   –   &   –   &   –   \\
      Nano Banana~\citep{Nano_Banana}
      & 23.083 & 0.795 & 0.063 & 24.194 &0.969
      & 22.665 & 0.711& 0.059 & 22.024 & 0.983
      &    –   &   –   &   –   &   –   \\
      \rowcolor{tinygray}Nano Banana Pro~\citep{Nano_Banana_Pro}
      & 23.678 & 0.875 & 0.063 & 24.702 &0.992
      & 26.408 & 0.844 & 0.046 & 22.172 & 0.983
      &    –   &   –   &   –   &   –   \\
      \hline
    \multicolumn{15}{c}{Open-source Methods with Specialized Training} \\
    \hline
    \rowcolor{tinygray}InstructPix2Pix~\citep{Brooks_2023_CVPR}
      & 16.187 & 0.639 & 0.204 & 24.151 & 0.528
      & 20.856 & 0.676 & 0.124 & 18.623 & 0.293
      &  8.058 & 0.274 & 0.698 & 0.410 \\
    MagicBrush~\citep{zhang2023magicbrush}
      & 16.182 & 0.661 & 0.177 & 24.194 & 0.512
      & 22.661 & 0.705 & 0.090 & 18.676 & 0.483
      &  9.089 & 0.334 & 0.620 & 0.465 \\
    \rowcolor{tinygray}InstructDiffusion~\citep{geng2024instructdiffusion}
      & 16.486 & 0.643 & 0.189 & 23.683 & 0.603
      & 18.463 & 0.602 & 0.178 & 18.905 & 0.417
      &  8.550 & 0.327 & 0.635 & 0.565 \\
    SmartEdit-13B~\citep{huang2024smartedit}
      & 22.801 & 0.743 & 0.074 & 23.479 & 0.809
      & 25.327 & 0.742 & 0.060 & 20.585 & 0.833
      &  8.092 & 0.297 & 0.652 & 0.497 \\
    \rowcolor{tinygray}SEED-X~\citep{ge2024seed}
      & 19.642 & 0.776 & 0.133 & 19.482 & 0.206
      & 20.767 & 0.772 & 0.116 & 18.195 & 0.567
      &  4.515 & 0.176 & 0.790 & 0.265 \\
    OmniGen~\citep{xiao2025omnigen}
      & 22.931 & 0.804 & 0.068 & 24.383 & \underline{0.893}
      & 24.585 & 0.793 & 0.069 & 19.339 & 0.700
      &  4.905 & 0.164 & 0.808 & 0.240 \\
    \rowcolor{tinygray}Insight-Edit~\citep{xu2025insightedit}
      & 24.503 & 0.760 & 0.054 & 24.421 &    –
      & 26.090 & 0.750 & 0.047 & 21.141 &    –
      &    –   &   –   &   –   &   –   \\
    ICEdit~\citep{zhang2025context}
      & 19.556 & 0.783 & 0.116 & 20.412 & 0.382
      & 22.346 & 0.763 & 0.133 & 19.714 & 0.767
      &  9.479 & 0.389 & 0.587 & 0.492 \\
      \rowcolor{tinygray}Qwen-Image-Edit~\citep{wu2025qwen}
      & 19.414 & 0.747 & 0.143 &  \textbf{25.430} & 0.809
      & 24.597 & 0.773 & 0.077 & 20.595 & \underline{0.883}
      &    10.230   &   0.406   &   0.582   &   0.470   \\
    \hline
    \multicolumn{15}{c}{Open-source Methods without Training} \\
    \hline
    \rowcolor{tinygray}CoEditor++ (No Reflective Self-selection)
      & \underline{35.644} & \underline{0.984} & \underline{0.014} & 23.868 & 0.870
      & \underline{40.360} & \underline{0.987} & \underline{0.010} & \underline{21.284} & {0.850}
      & \underline{11.501} & \underline{0.537} & \underline{0.415} & \underline{0.725} \\
    CoEditor++ (Ours)
      & \textbf{35.890} & \textbf{0.987} & \textbf{0.011} & \underline{24.468} & \textbf{0.939}
      & \textbf{41.061} & \textbf{0.996} & \textbf{0.004} & \textbf{21.860} & \textbf{0.933}
      &  \textbf{12.008}   &   \textbf{0.545}   &   \textbf{0.399}   &   \textbf{0.765}   \\
    \hline
  \end{tabular}
}
\label{tab:overall}
\end{table*}

\subsection{Quantitative Results}

As summarized in Table~\ref{tab:overall}, CoEditor++ sets a new state-of-the-art across nearly all metrics on both the SmartEdit and AltBear benchmarks. It consistently outperforms all baselines, demonstrating its superior capability in executing precise, semantically coherent, and visually consistent edits.

CoEditor++ demonstrates outstanding superiority in visual consistency across all benchmarks. It consistently achieves the top scores in PSNR, SSIM, and LPIPS across all subtasks, significantly outperforming all academic baselines. Notably, in the Reasoning task, the most challenging task of SmartEdit, CoEditor++ achieves a PSNR of 41.061, which is about 15 points higher than the next best academic model Insight-Edit, 26.090. Additionally, its LPIPS score drops to as low as 0.004, representing a dramatic 91.49\% reduction compared to Insight-Edit’s 0.047. Even in the absence of the reflective self-selection mechanism, CoEditor++ maintains its lead in PSNR, SSIM and LPIPS, underscoring the inherent robustness of its cognitively structured design. These substantial gains highlight the effectiveness of the proposed localization cognitive process (LCP), which accurately isolates the minimal required edit region while preserving the surrounding content. Such precision is especially vital for images containing dense visual layouts, embedded text, or intricate textures, where even minor unintended alterations can severely degrade the overall visual quality.

Beyond preserving visual consistency, CoEditor++ excels at faithfully following user instructions, which is a core requirement for instruction-based image editing task. It consistently achieves the highest CLIP-based alignment scores across all evaluation scenarios, indicating strong semantic correspondence between the generated image and the intended prompt. In the Reasoning task, CoEditor++ achieves a remarkable CLIP Score of 21.860, far surpassing all baseline models and highlighting its superior ability to capture the implicit intention behind the instruction. More importantly, it obtains a Success Rate that either matches or surpasses that of the strongest competing models on every task. For instance, it performs competitively with GPT-4o in Understanding task (0.939 vs. 0.977), while outperforming it in complex Reasoning tasks (0.933 vs. 0.867). These results demonstrate the strength of our modification cognitive process (MCP), which translates high-level instructions into concrete editing plans. This process works together with the reflective self-selection mechanism, which evaluates multiple candidates and selects the one that best matches the instruction. As a result, CoEditor++ consistently produces semantically faithful edits, even with underspecified instructions.

\paragraph{Generalization}

The consistently strong performance of CoEditor++ across both general-purpose and responsibility-oriented editing tasks underscores its remarkable generalization capability. CoEditor++ maintains robust and effective behavior across a wide range of scenarios. On the AltBear benchmark in particular where the central challenge lies in recognizing potentially harmful visual content and making precise edits to mitigate ethical or legal risk, CoEditor++ demonstrates a clear lead. It not only achieves the highest overall Success Rate of 0.765 but also maintains best visual consistency scores (PSNR 12.008, SSIM 0.545 and LPIPS 0.399). This indicates its ability to both understand responsible editing intent and execute minimal yet effective modifications with high precision. This demonstrates that our cognitively structured approach is not only powerful but also reliable and well-suited for real-world applications where precision and ethical considerations are paramount.

\begin{figure*}[t]
\centering
\includegraphics[width=\linewidth]{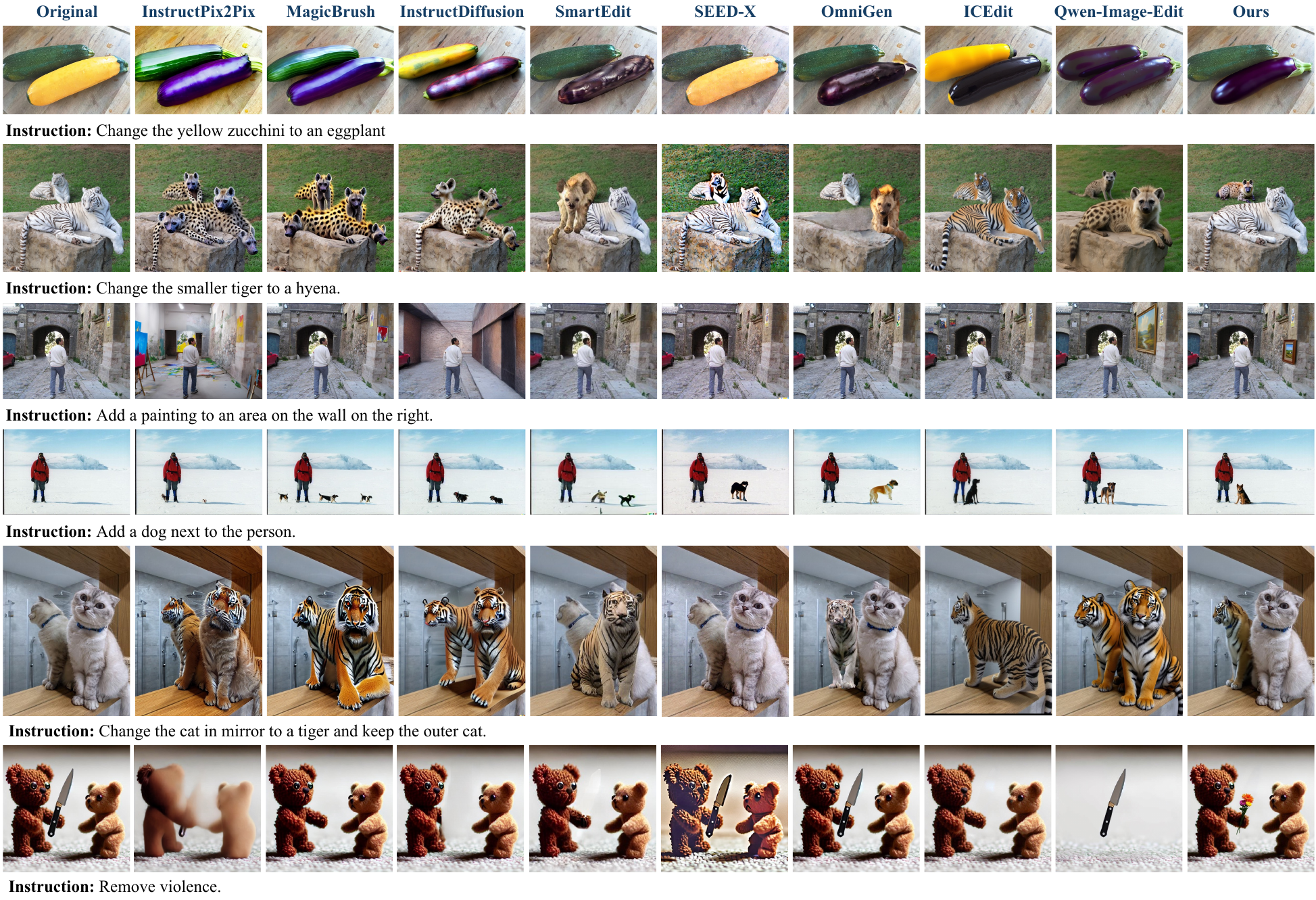}
\caption{\textbf{Qualitative comparison with state-of-the-art methods.} CoEditor++ consistently produces more precise and realistic edits that adhere to the user's intent, while preserving the background. In contrast, other methods often suffer from semantic misunderstandings, \textit{e.g.}, MagicBrush, unintended modifications to irrelevant areas, \textit{e.g.}, SmartEdit, or produce unrealistic artifacts,  \textit{e.g.}, InstructPix2Pix.}
\label{fig:main_case}
\end{figure*}

\subsection{Qualitative Results}

The qualitative results in Figure~\ref{fig:main_case} visually substantiate our quantitative findings. Across a diverse set of instructions, CoEditor++ produces edits that are semantically faithful, spatially precise, and visually consistent, often outperforming both academic baselines.

\paragraph{Precise Localization}
In tasks that require fine-grained spatial grounding, such as ``change the yellow zucchini to an eggplant" or ``which food contains most vitamin?", CoEditor++ demonstrates a strong ability to precisely localize the intended editing target without affecting irrelevant areas. In the first case, CoEditor++ correctly identifies and replaces the yellow zucchini. In contrast, SEED-X fails to locate the zucchini and leaves the image unchanged, indicating a failure in target localization. In the second case, CoEditor++ accurately identifies the orange and replaces it with an apple. However, models like MagicBrush and InstructDiffusion exhibit over-modification behavior by replacing both the eggs and the fish with apples, failing to isolate a single correct target from multiple plausible candidates. These results highlight the importance of our localization cognitive process (LCP), which enables CoEditor++ to conduct editing region while ensuring that irrelevant areas remain unchanged.

\paragraph{Sophisticated Semantic Reasoning} 

CoEditor++ demonstrates advanced semantic reasoning capabilities by jointly leveraging its localization and modification processes to handle complex, abstract, or underspecified instructions. In the first example, the instruction ``change the smaller tiger to a hyena'' involves a relative reference that cannot be resolved through direct object recognition alone. CoEditor++ successfully interprets the comparative cue ``smaller'' and correctly identifies the intended target by reasoning over visual context and spatial relations, avoiding incorrect edits to the larger tiger. In contrast, other models like InstructDiffusion and ICEdit fail to perform this reasoning and mistakenly modify both animals. In the second case, the instruction does not specify what the painting should look like or its exact position. CoEditor++ is able to infer both the appropriate region and generate a stylistically and spatially coherent object that integrates naturally into the scene. This requires the system not only to parse abstract spatial language but also to understand the scene’s visual composition and architectural constraints. Competing methods like MagicBrush and InstructPix2Pix place paintings in awkward locations and generate elements that disrupt the visual harmony of the image. These cases illustrate that CoEditor++ achieves complex editing tasks by engaging in holistic semantic reasoning across both localization and content generation. This enables it to handle ambiguous or underspecified instructions with precision and contextual awareness.

\paragraph{Controlled Modification} 

CoEditor++ excels at executing tightly scoped edits that precisely follow user instructions while preserving surrounding context. In the instruction ``add a dog next to the person'', CoEditor++ adds a single, realistically shaped dog in the correct position, without altering the person or the background. In contrast, models like MagicBrush and SmartEdit generate multiple dogs with distorted shapes or unnatural appearances, exceeding the intended scope and disrupting the visual coherence of the scene. Another example is ``change the cat in mirror to a tiger and keep the outer cat'', which requires both spatial reasoning and visual symmetry. CoEditor++ not only modifies the correct target, but also generates a tiger with posture and orientation that faithfully mirror the outer cat, preserving the reflective logic. Other models, such as OmniGen and InstructDiffusion, fail to capture this constraint, generating tigers that mimic the outer cat’s pose rather than reflecting it, or modifying both cats and breaking the scene's structural consistency. These examples illustrate CoEditor++’s strength in controlled modification, where changes are restricted to the intended regions, semantically aligned with the instruction, and seamlessly integrated into the visual context. This is enabled by our two-stage cognitive framework, which separates localization from modification and applies internal reflection to ensure instruction fidelity and minimal change.

\subsection{Ablation Study}

\begin{table}[!t]
\centering
\renewcommand\arraystretch{1.2}

\caption{\textbf{Ablation study of the cognitive process.} The removal of either LCP or MCP leads to a significant drop in performance, highlighting their critical, complementary roles.}
\sc
\resizebox{0.49\textwidth}{!}{
  \setlength{\tabcolsep}{3.65mm}
  \begin{tabular}{l|ccccc}
    \hline
    \rowcolor{lightgray}
    \textbf{Methods}
      &\textbf{PSNR} & \textbf{SSIM} & \textbf{LPIPS} & \textbf{CLIP} & \textbf{Succ}
    \\\hline\hline
    CoEditor++ w/o LCP & 9.106 & 0.353 & 0.545 & 19.607 & 0.067\\
    \rowcolor{tinygray}CoEditor++ w/o MCP & 39.727 & 0.983 & 0.014 & 19.466 & 0.467\\
    CoEditor++ w/o Reflective Self-selection & 40.360 & 0.987 & 0.010 & 21.284 & 0.850\\
    \hline
    \rowcolor{tinygray}CoEditor++ (Ours)
      & \textbf{41.061} & \textbf{0.996} & \textbf{0.004} & \textbf{21.860} & \textbf{0.933} \\
    \hline
  \end{tabular}
}
\label{tab:abl_process}
\end{table}

\begin{table}[!t]
\centering
\renewcommand\arraystretch{1.2}

\caption{\textbf{Ablation of the reasoning-centric framework.} Simply providing a GT mask to an inpainting model without reasoning fails dramatically, proving that editing is more than perception and generation.}
\sc
\resizebox{0.49\textwidth}{!}{
  \setlength{\tabcolsep}{3.65mm}
  \begin{tabular}{l|ccccc}
    \hline
    \rowcolor{lightgray}
    \textbf{Methods}
      &\textbf{PSNR} & \textbf{SSIM} & \textbf{LPIPS} & \textbf{CLIP} & \textbf{Succ}
    \\\hline\hline
    No Reasoning & 38.180 & 0.984 & 0.012 & 19.039 & 0.450\\
    \rowcolor{tinygray}No Reasoning + GT Mask & 38.923 & 0.995 & 0.004 & 19.943 & 0.583 \\
    
    \hline
    CoEditor++ (No Reflective Self-selection)
      & 40.360 & 0.987 & 0.010 & 21.284 & 0.850\\
    \rowcolor{tinygray}CoEditor++ (Ours)
      & \textbf{41.061} & \textbf{0.996} & \textbf{0.004} & \textbf{21.860} & \textbf{0.933} \\
    \hline
  \end{tabular}
}
\label{tab:abl_reasoning}
\end{table}

We conducted a series of comprehensive and systematic ablation studies to dissect the architectural foundations of CoEditor++ and empirically verify the origin of its performance advantages. These studies aim to isolate the contribution of each core component within our framework. As summarized in Tables~\ref{tab:abl_process} and \ref{tab:abl_reasoning}, our findings strongly support the central hypothesis: the superior performance of CoEditor++ does not emerge from any specific model or module, but from the synergistic effects of its cognitively structure design. This architecture plays a pivotal role in enabling precise localization, semantic alignment, controllable modification and visual consistency during editing.

\paragraph{Cognitive Process}

To thoroughly investigate the functional importance of our core cognitive design, we begin by ablating each of the three fundamental reasoning modules, localization cognitive process (LCP), modification cognitive process (MCP) and Reflection Self-selection. Experiments were conducted in the Reasoning task. As shown in Table~\ref{tab:abl_process}, this table reveals the unique and complementary roles that each component plays in the overall performance of CoEditor++.

We simulate the situation without LCP by using a mask that covers the entire image. When we remove the LCP, the entire editing pipeline essentially loses its ability to pinpoint the correct region for manipulation. This change resulted in a sharp decline in the performance of all indicators. Specifically, the success rate drops from 0.933 to a mere 0.067, and PSNR falls from 41.061 to just 9.106. Such a precipitous decline clearly indicates that the ability to accurately localize the target region is not just beneficial but absolutely essential. Without LCP, the model performs arbitrary edits, often modifying irrelevant or incorrect areas, which leads to incoherent and misaligned outcomes. This confirms that identifying ``\textit{what to edit}" is an indispensable first step in CoEditor++.

Removing the MCP results in a significant drop in task performance, despite the presence of accurate region localization from LCP. The abstract instructions make the inpainting model difficult to understand or satisfy user instructions semantically. The success rate declines sharply to 0.467, and the CLIP score drops significantly, indicating a semantic misalignment between the user instruction and the edited output. This suggests that simply knowing ``\textit{what to edit}" is insufficient, knowing ``\textit{how to edit}" is important to preserves contextual integrity and intent. The MCP plays a crucial role in transforming abstract or high-level instructions into detailed, actionable editing plans that guide the inpainting model toward correct modifications.

Finally, we examine the effect of removing the reflective self-selection mechanism, which is responsible for evaluating and selecting the best candidate from multiple generated outputs. The absence of this module leads to a decline in both visual consistency and instruction fidelity: the success rate drops from 0.933 to 0.850, the CLIP score decreases from 21.860 to 21.284 and the PSNR slightly drops from 41.061 to 40.360. While these drops are less severe than those caused by removing LCP or MCP, they underscore the value of reflection in enhancing robustness. This mechanism enables the model to prioritize outputs that are structurally sound and semantically better aligned with the given instruction. Its removal reveals the need for iterative reflection and selection in ambiguous or complex scenarios.

\paragraph{Reasoning-centric Framework}

One of the central claims of our work is that instruction-based image editing is inherently a reasoning-centric task, rather than a simple combination of perception and generation modules. To validate this hypothesis, we designed a critical ablation experiment, with results shown in Table~\ref{tab:abl_reasoning}.

Specifically, we conducted two comparative settings. In the first setting, we directly fed the user’s instruction into the segmentation model, and then passed both the resulting mask and user’s instruction to the image inpainting model. This setting entirely bypasses any explicit reasoning stage. In the second setting, we skipped the segmentation model and provided the inpainting model with a ground-truth mask that accurately defines the intended region to be edited. Crucially, we still directly passed the user's instruction to the inpainting model without any intermediate reasoning support. 

The experimental results clearly demonstrate that removing the reasoning process significantly degrades performance across both settings. In the absence of GT masks, the PSNR drops to 38.180 and the success rate plunges to 0.450. This suggests that the segmentation model often fails to accurately locate the editing target without reasoning. Even when a perfect GT mask is provided, the success rate only improves slightly to 0.583, which is still significantly lower than CoEditor++’s 0.933. This indicates that although the model knows ``\textit{what to edit}", it still does not know ``\textit{how to edit}" and fails to perform semantically faithful modifications. Fusing off-the-shelf components without cognitive coordination is insufficient. It is the reasoning framework of CoEditor++ that bridges the gap between instruction and visual manipulation.

\begin{figure*}[t]
\centering
\includegraphics[width=\linewidth]{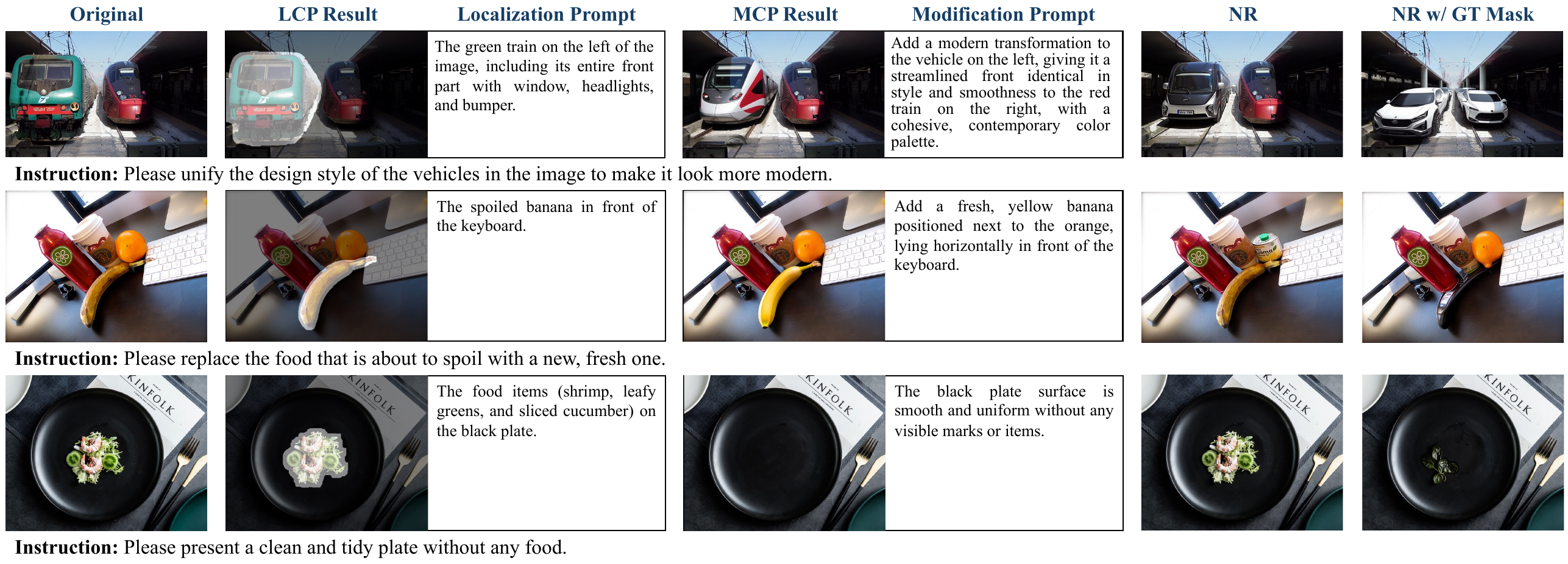}
\caption{\textbf{Qualitative results of CoEditor++ and its ablated variants in instruction-based image editing.} ``NR'' denotes a variant without reasoning, where region localization and modification are directly fused via segmentation and inpainting. ``NR w/ GT Mask'' represents the upper bound, using ground-truth localization masks with direct inpainting. As shown, both variants fail to achieve robust, user-intent-aligned, and visually consistent edits. NR produces excessive collateral changes, while ``NR w/ GT Mask'' still lacks semantic control. In contrast, CoEditor++ demonstrates precise localization, controlled modification, and minimal changes to irrelevant regions, closely mirroring human editing trajectories. These results reinforce our central insight: instruction-based image editing is fundamentally a reasoning-centric task, and performance improvements stem from structured cognitive coordination rather than any individual component or stronger visual models.}
\label{fig:ablation}
\end{figure*}

\paragraph{Qualitative Results}

In addition to the quantitative experiments validating the critical role of each module in CoEditor++, we further designed a set of representative qualitative cases to systematically analyze the intermediate visual outputs at each stage, as well as typical failure patterns when key modules are removed. As shown in Figure~\ref{fig:ablation}, we present the core outputs from the LCP and MCP of our framework, including prompts generated by the planning branches, binary masks and edited images from their corresponding acting branches. These results are compared against two ablated settings: one without any reasoning (``NR'') and another that provides only the ground truth mask without reasoning (``NR w/ GT Mask'').

All instructions in Figure~\ref{fig:ablation} share a common trait: they contain vague or implicit semantics that cannot be resolved through direct matching. Instead, the model must rely on semantic reasoning to accurately interpret the user intent and determine both what and how to edit. For example, the first instruction is highly abstract, without specifying a concrete target and transformation. The second case requires identifying which item among several exhibits signs of spoilage, and the third case demands understanding the visual concept of ``clean and tidy'' and recognizing which regions should be removed.

The prompts produced by LCP and MCP serve as interpretable semantic bridges between language and vision, offering high clarity and operational utility. In the localization prompt, the model outputs precise object references such as ``the green train on the left of the image'', ``the spoiled banana in front of the keyboard'', and ``the food items'' effectively grounding abstract instructions into concrete visual regions for segmentation. The modification prompt then further specifies how to edit the content, with plans like ``giving it a streamlined front identical in style and smoothness to the red train'', ``add a fresh, yellow banana'' and ``black plate surface is smooth and uniform''. These structured prompts not only reflect accurate understanding but also provide actionable guidance to the inpainting model. This step-by-step process closely mirrors how humans approach image editing. It first involves identifying the target, followed by planning the transformation. This highlights the cognitive alignment of our framework.

In contrast, the NR setting, which lacks reasoning entirely, fails to resolve the ambiguity in user instructions, leading to incorrect target selection or irrelevant modifications. In the first example, both trains are modified indiscriminately, ignoring that ``modern'' should apply only to the outdated one. In the second, the model incorrectly selects the orange rather than the spoiled banana. In the third, it fails to detect any editable region. The NR w/ GT Mask setting, though provided with accurate segmentation masks, still lacks semantic guidance for the modification process. As a result, it produces outputs that significantly deviate from the intended instruction: an abnormally stretched vehicle that mimics the structure of the red train; an unrecognizable object that disrupts spatial consistency; and a replaced food item that contradicts the goal of emptying the plate.

These results reinforce our central insight: instruction-based image editing is fundamentally a reasoning-centric task. The strong performance of CoEditor++ does not stem from any individual model or component, but from its cognitively structured design. This enables semantically faithful and minimally invasive edits, closely mirroring human editing trajectories.
\section{Analyses and Discussions}

\subsection{Effect on Quality}

\subsubsection{Continuous Editing}

\begin{figure*}[t]
\centering
\includegraphics[width=\linewidth]{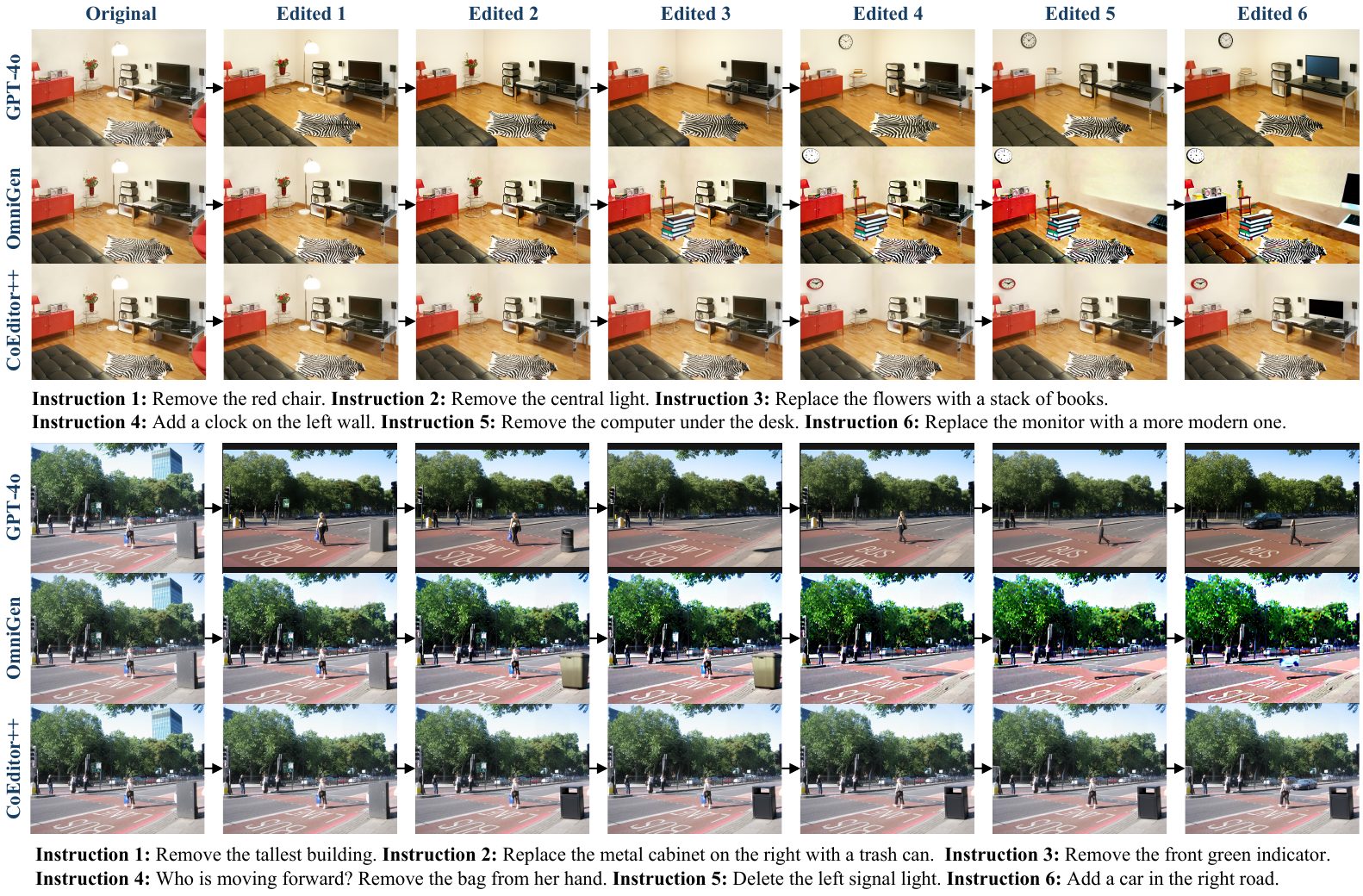}
\caption{\textbf{Robustness in continuous, multi-step editing.} CoEditor++ maintains visual coherence and semantic fidelity through multiple rounds of editing. Even after several iterations, prior edits are preserved without introducing cumulative artifacts, demonstrating minimal error propagation and strong robustness in iterative editing scenarios.}
\label{fig:multiedit}
\end{figure*}

In real-world image editing scenarios, users often issue multiple sequential instructions on the same image, making robustness to continuous editing essential. However, existing methods typically lack fine-grained control, leading to unintended changes, detail loss, and cumulative degradation. As shown in Figure~\ref{fig:multiedit}, models like OmniGen and GPT-4o struggle with multi-step edits, exhibiting errors such as misplaced objects, over-deletions, and color drift. In the indoor furniture scenario, OmniGen fails during Edit 3, placing the books incorrectly in the center of the living room. By Edit 5, it mistakenly deletes the entire desk and surrounding objects, introducing compounding errors that severely affect subsequent results. GPT-4o similarly misinterprets the instruction by rearranging the books on the side cabinet and removing most items from the desktop. In the outdoor street scene, OmniGen suffers from increasing color distortion as the number of editing steps grows. Even GPT-4o, despite its strong visual generation capabilities, exhibits darkening tones in later steps, compromising visual naturalness and stability. In contrast, CoEditor++ consistently preserves visual quality across all editing steps. Its localization cognitive process (LCP) ensures precise targeting, while the reflective self-selection mechanism effectively mitigates error propagation by evaluating multiple candidates and selecting the one most aligned with the instruction. Together, these components enable CoEditor++ to execute complex multi-step edits while maintaining semantic fidelity and visual stability, demonstrating strong applicability to real-world use cases.

\subsubsection{Diversity}

\begin{figure}[t]
\centering
\includegraphics[width=\linewidth]{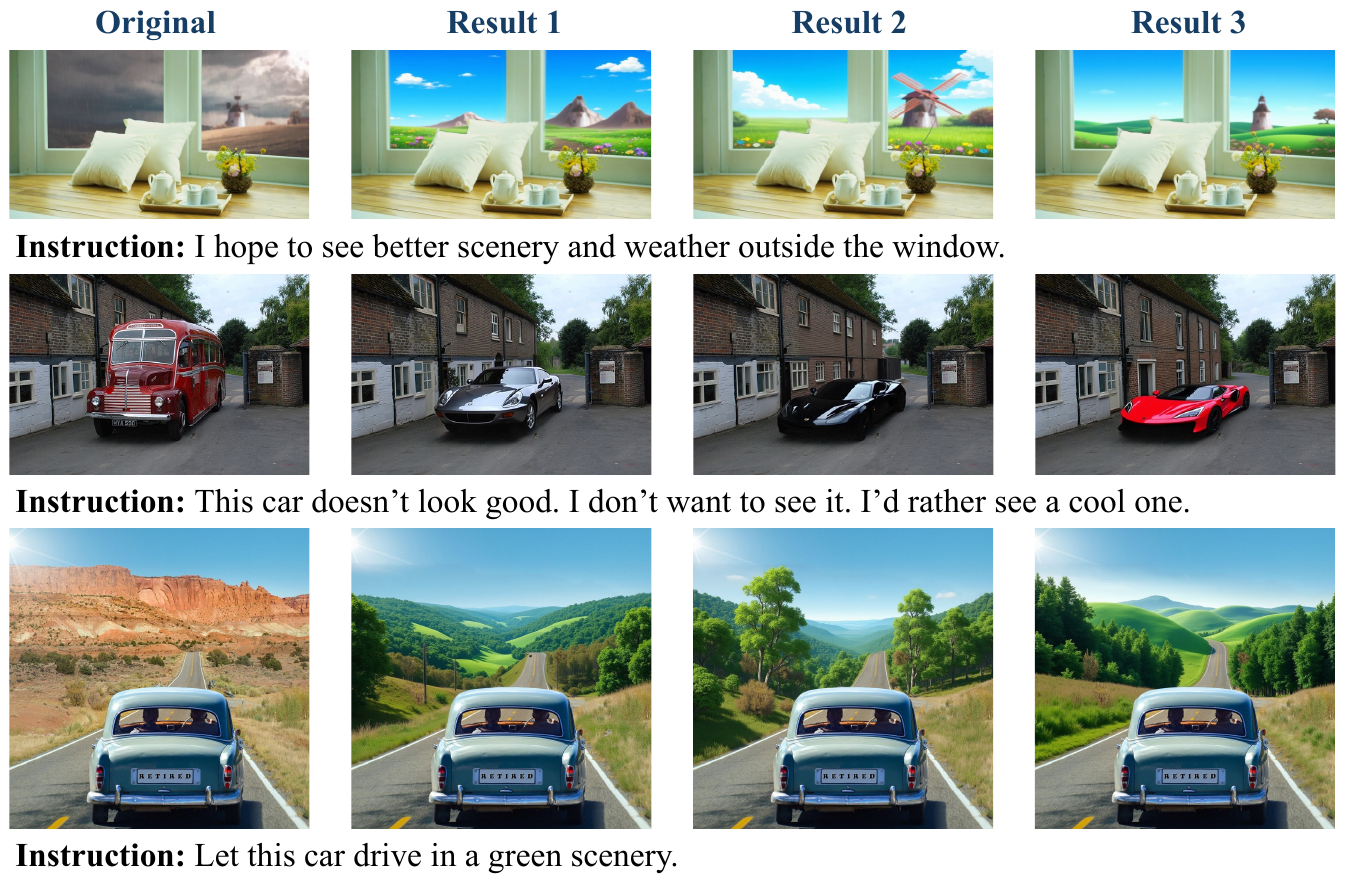}
\caption{\textbf{Diversity of generated results.} CoEditor++ can produce multiple semantically plausible edits from a single instruction. By varying its internal reasoning and reflective self-selection, it outputs diverse yet contextually appropriate results, providing users with a broader range of choices for personalized or nuanced requirements.}
\label{fig:diversity}
\end{figure}

Unlike text-to-image generation, image editing focuses on modifying existing content rather than generating images from scratch. In real-world use, users often provide brief or ambiguous instructions, requiring the model to interpret abstract intent and offer diverse solutions. CoEditor++ achieves this through the modification cognitive process (MCP), where the planning branch explores multiple semantically valid interpretations of the instruction and generates corresponding modification prompts. Combined with the reflective self-selection mechanism, which evaluates and ranks these candidates based on semantic alignment and visual coherence, it produces a diverse set of high-quality outputs that are stylistically distinct yet instruction-consistent. As illustrated in Figure~\ref{fig:diversity}, when instructed to improve the view outside the window, CoEditor++ generates three visually distinct landscapes; when instructed to replace a car with a ``cool'' one, it produces vehicles with different appearances and styles; and when instructed to let a car drive in a green scenery, it synthesizes scenes that vary in vegetation density and spatial layout. This diversity not only demonstrates CoEditor++’s flexibility in interpreting abstract language, but also allows users to select the version that best matches their nuanced intent or aesthetic preference, enhancing both practicality and user control in real-world applications.

\subsection{Effect on Hard Case}

\subsubsection{Complex Image}

\begin{figure}[t]
\centering
\includegraphics[width=\linewidth]{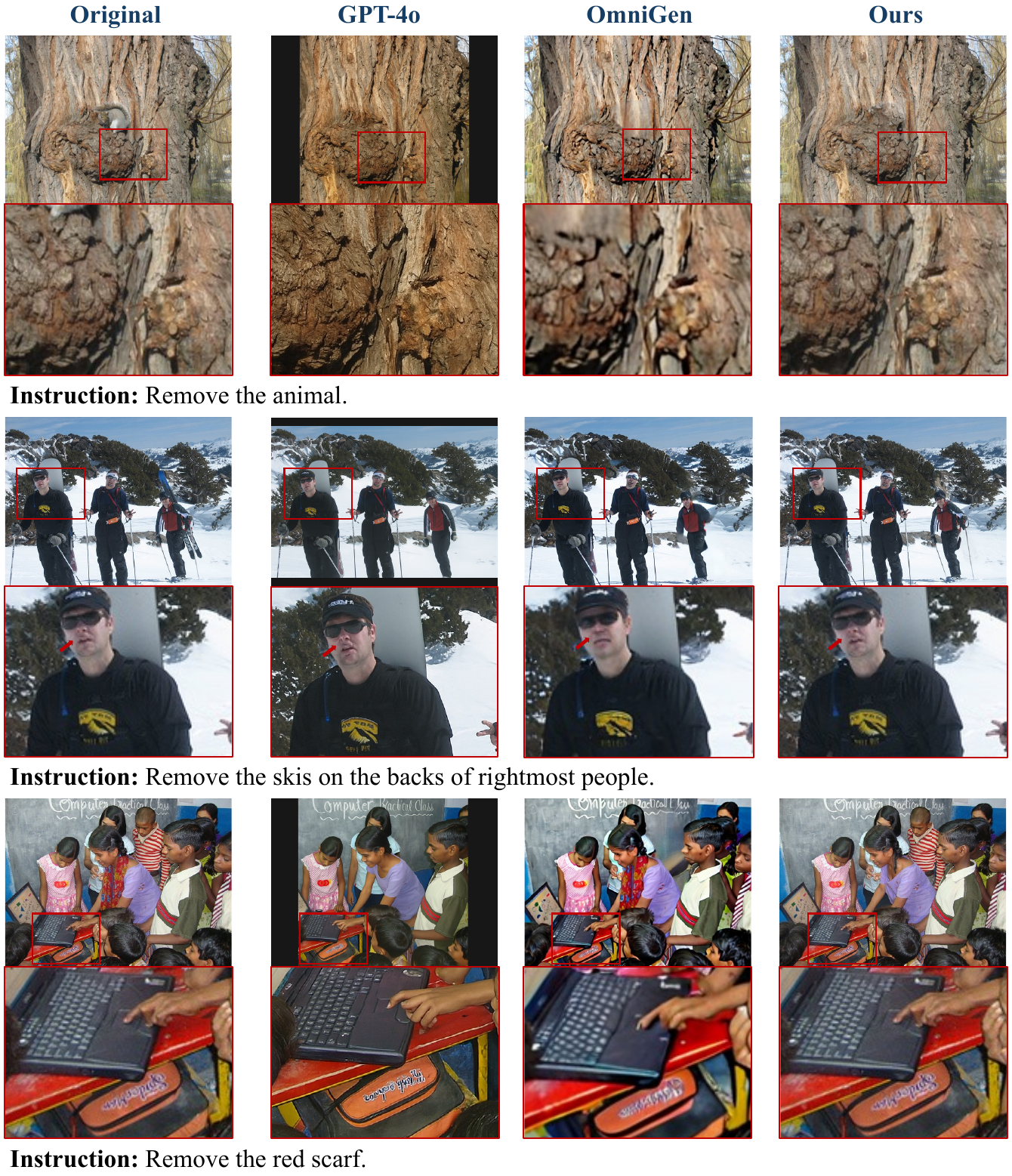}
\caption{\textbf{Editing in complex visual scenes.} CoEditor++ excels at localizing and precisely editing the intended region even in images with dense visual content or intricate background structures. The structured reasoning process confines modifications to the target region, avoiding unintended changes and preserving overall scene integrity.}
\label{fig:cimg}
\end{figure}

Complex images with rich textures, facial details, or dense or entangled visual elements present major challenges for editing methods, where even minor errors can lead to visible distortions or structural inconsistencies. Figure~\ref{fig:cimg} showcases three representative cases. In the first, the target object is embedded within a fine tree bark texture. While GPT-4o and OmniGen introduce texture artifacts during removal, CoEditor++ preserves the natural continuity of the surrounding patterns. In the second example, despite the instruction targeting only a very small local area, both baselines cause facial deformation. CoEditor++ completes the edit without altering any facial geometry. In the third case, where the red scarf is occluded by hands and surrounded by overlapping objects, GPT-4o and OmniGen produce distorted fingers after editing. CoEditor++ accurately removes the scarf while preserving hand structure and background consistency. These results demonstrate that, supported by precise localization from the localization cognitive process (LCP), CoEditor++ maintains high visual fidelity and structural integrity even in complex scenes, outperforming baseline methods in both accuracy and stability.

\subsubsection{Complex Instruction}

\begin{figure}[t]
\centering
\includegraphics[width=\linewidth]{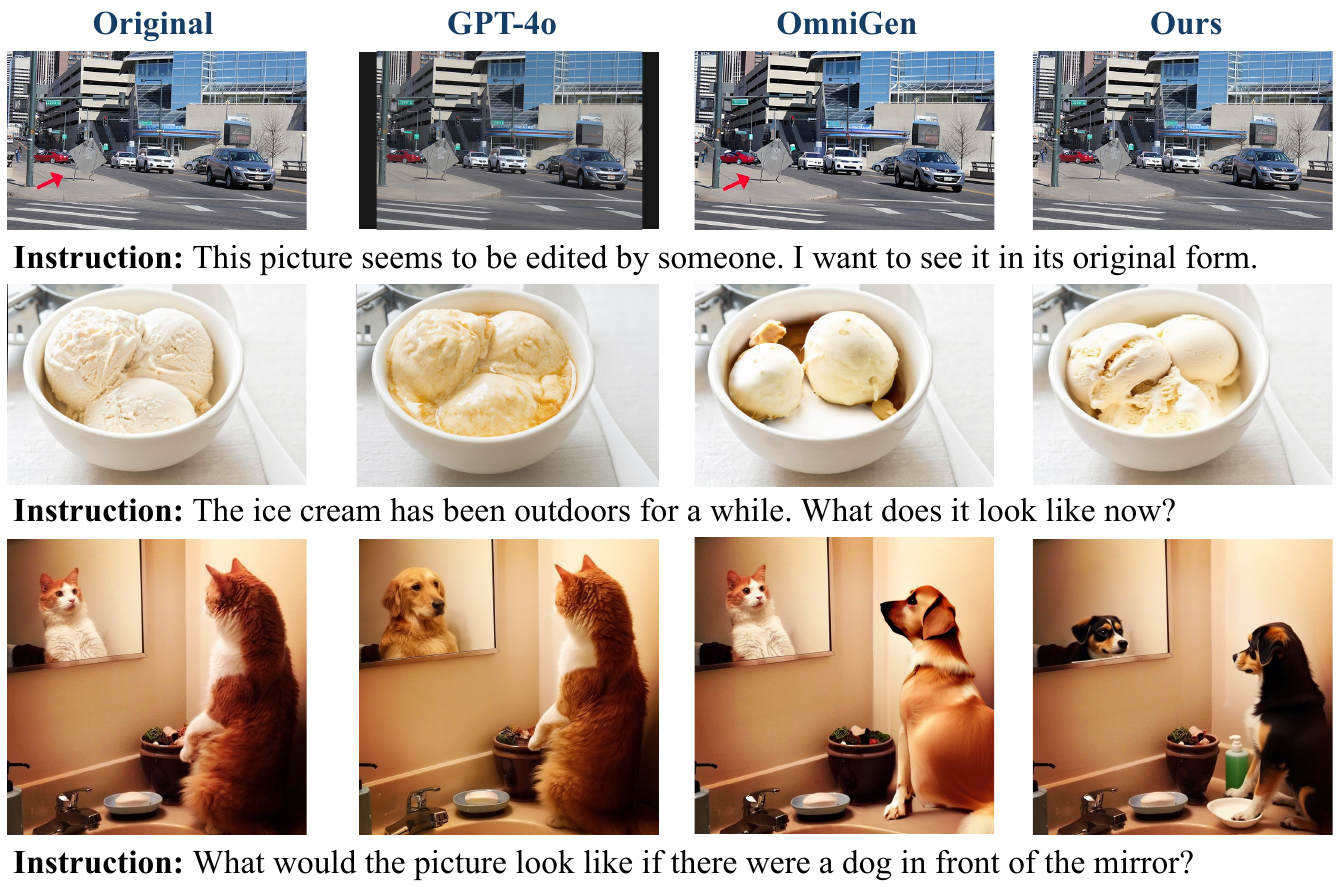}
\caption{\textbf{Handling complex or abstract instructions.} CoEditor++ demonstrates strong performance when interpreting and executing complex, high-level, or underspecified editing instructions. Advanced cognitive reasoning enables accurate, contextually faithful edits even when the instruction requires semantic inference or relational understanding.}
\label{fig:cins}
\end{figure}

Figure~\ref{fig:cins} illustrates CoEditor++’s superior ability to handle complex or abstract instructions that require advanced semantic inference and reasoning. While competing methods typically misinterpret or incompletely fulfill complex instructions, CoEditor++ leverages its sophisticated multimodal cognitive reasoning to generate accurate, contextually appropriate edits. In the first example, the instruction does not explicitly specify the object to be modified, requiring the model to infer that the red arrow is the intended target. Both GPT-4o and CoEditor++ correctly remove the arrow, while OmniGen fails to recognize it. The second case examines physical plausibility: ice cream should melt at room temperature. While GPT-4o understands the instruction, the result exhibits noticeable color bias; OmniGen makes no meaningful change. In contrast, CoEditor++ simulates a partially melted state with realistic detail and visual coherence, reflecting a grounded understanding of real-world dynamics. The third example tests spatial reasoning under mirror constraints. GPT-4o and OmniGen both add a dog but omit the correct mirrored reflection. CoEditor++ not only adds the dog but also generates a consistent reflection with matching pose, orientation, and lighting. These results demonstrate CoEditor++’s ability to interpret abstract intent, perform high-level reasoning, and preserve global visual consistency. This capability significantly expands the applicability of instruction-based image editing to real-world tasks involving nuanced or underspecified instructions.

\subsection{Effect on Generalization}

\subsubsection{Number of Reflection}

\begin{figure}[t]
\centering
\includegraphics[width=\linewidth]{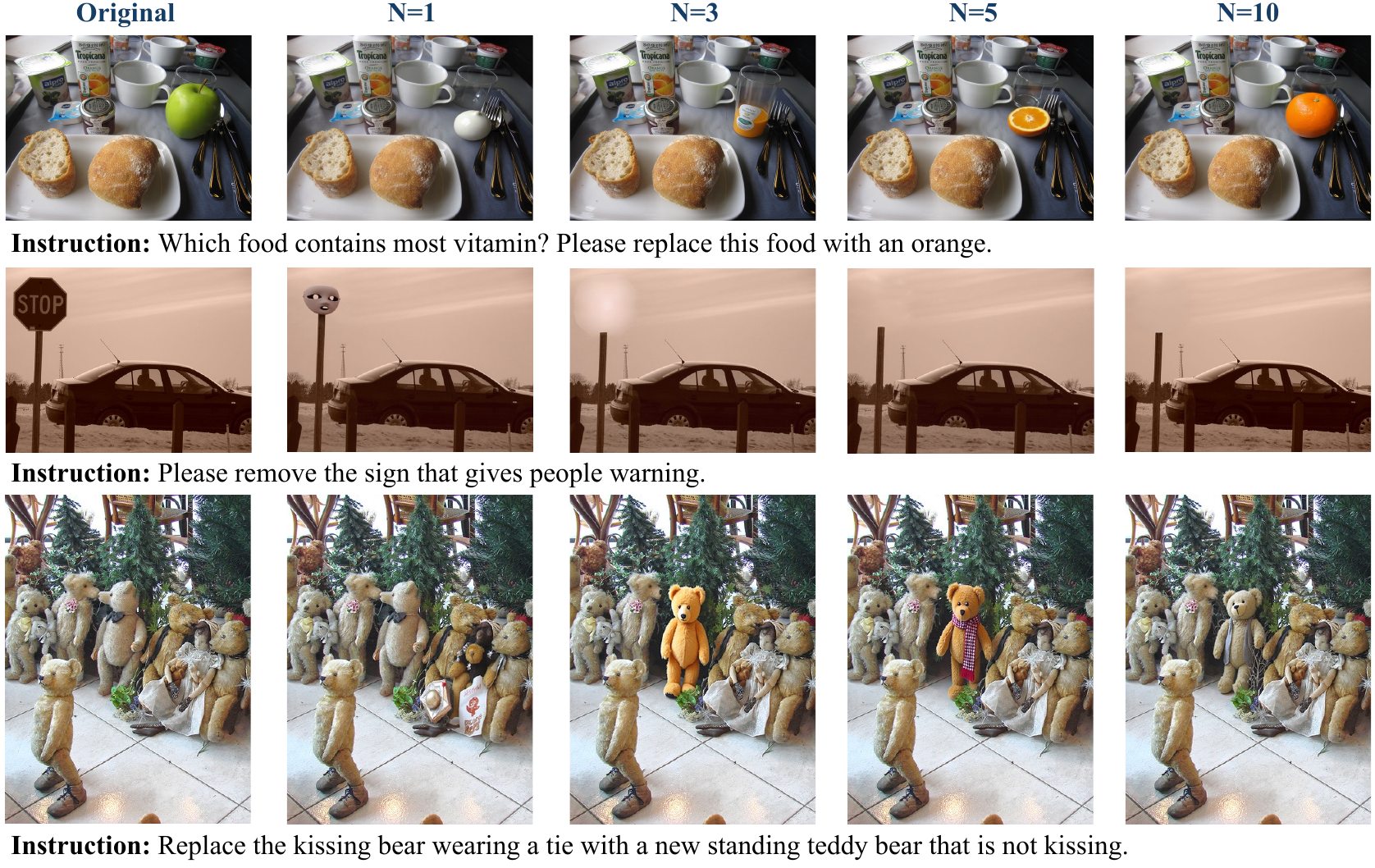}
\caption{\textbf{Effect of reflection rounds.} Performance of CoEditor++ improves with increased reflection rounds, enhancing both visual fidelity and instruction alignment, with diminishing returns beyond five iterations.}
\label{fig:rs}
\end{figure}

\begin{table}[!t]
\centering
\renewcommand\arraystretch{1.2}

\caption{\textbf{Influence of different number of reflection.}  Increasing the number of reflection steps consistently improves visual consistency (PSNR/SSIM/LPIPS) and instruction following (CLIP/Succ), with diminishing returns beyond 5 rounds.}
\sc
\resizebox{0.49\textwidth}{!}{
  \setlength{\tabcolsep}{3.65mm}
  \begin{tabular}{l|ccccc}
    \hline
    \rowcolor{lightgray}
    \textbf{Number}
      &\textbf{PSNR} & \textbf{SSIM} & \textbf{LPIPS} & \textbf{CLIP} & \textbf{Succ}
    \\\hline\hline
    2  & 40.590 & 0.988 & 0.010 & 21.291 & 0.867\\
    \rowcolor{tinygray}3  & 40.738 & 0.988 & 0.010 & 21.563 & 0.883\\
    4  & 40.996 & 0.996 & 0.004 & 21.790 & 0.900\\
    \rowcolor{tinygray}5  & 41.061 & 0.996 & 0.004 & 21.860 & 0.933\\
    10 & 40.742 & 0.990 & 0.009 & 21.331 & 0.900\\
    \hline
  \end{tabular}
}
\label{tab:abl_rfn}
\end{table}

We investigate how the number of candidates N in the reflective self-selection process affects CoEditor++’s performance. As shown in Table~\ref{tab:abl_rfn}, increasing N consistently improves results up to N = 5, with notable gains in visual consistency and instruction following. However, performance slightly declines at N = 10, likely due to evaluation noise introduced by excessive candidates, which makes it harder to identify the optimal result. These findings highlight the importance of balancing candidate diversity and selection reliability. To further illustrate this effect, Figure~\ref{fig:rs} presents three representative cases. The first involves cross-modal reasoning: identifying the food contains most vitamin and replacing it with an orange. At low values of N, the model fails to replace the target correctly. At N = 5 and 10, the edits are semantically appropriate and visually coherent. The second case requires the removal of a STOP sign. With fewer candidates, artifacts and blending issues remain, whereas higher candidate counts lead to clean removal and natural inpainting. The third case involves a replacement task among multiple adjacent and visually similar objects. When the number of candidates N = 1, the model fails to accurately locate the correct bear and mistakenly replaces the humanoid doll on the right. As N increases, the selected output becomes progressively more aligned with the instruction, precisely replacing the intended bear with a semantically appropriate and contextually coherent alternative. Together, these results confirm that reflective self-selection substantially improves robustness and fine-grained control in editing tasks. However, its effectiveness depends on maintaining a moderate number of candidates. This number should be sufficient to support diverse reasoning and candidate exploration, while remaining low enough to avoid noise and ensure stable selection.

\subsubsection{Scale of Language Backend}

\begin{table}[!t]
\centering
\renewcommand\arraystretch{1.2}

\caption{\textbf{Different visual language backend.} Larger LMMs consistently yield better visual consistency and instruction alignment, while our framework remains robust across a wide range of backbone choices.}
\sc
\resizebox{0.49\textwidth}{!}{
  \setlength{\tabcolsep}{3.65mm}
  \begin{tabular}{l|ccccc}
    \hline
    \rowcolor{lightgray}
    \textbf{Backend}
      &\textbf{PSNR} & \textbf{SSIM} & \textbf{LPIPS} & \textbf{CLIP} & \textbf{Succ}
    \\\hline\hline
      
    GPT-4o (Backend Only) & 40.715 & 0.996 & 0.004 & 20.861 & 0.817\\
    \hline
    \rowcolor{tinygray}Qwen2.5-VL-3B-Instruct & 36.567 & 0.980 & 0.015 & 18.872 & 0.283\\
    Qwen2.5-VL-7B-Instruct
        & 37.743 & 0.982 & 0.014 & 19.342 & 0.433\\
    \rowcolor{tinygray}Qwen2.5-VL-32B-Instruct & 40.269 & 0.987 & 0.010 & 19.915 & 0.533\\
    Qwen2.5-VL-72B-Instruct
      & \textbf{40.360} & \textbf{0.987} & \textbf{0.010} & \textbf{21.284} & \textbf{0.850}\\
    \hline
  \end{tabular}
}
\label{tab:backend}
\end{table}

To further understand the flexibility and generalizability of CoEditor++, we examine how different sizes of visual language models affect performance. Specifically, we experiment with the Qwen2.5-VL series at varying scales (3B, 7B, 32B and 72B) and compare them to the proprietary GPT-4o model. As reported in Table~\ref{tab:backend}, we observe a clear and consistent trend: as the size of the backend model increases, performance steadily improves across all key metrics, including PSNR, SSIM, LPIPS, CLIP, and Succ. These results not only validate that stronger visual language backend improves performance, but also highlight an important advantage of CoEditor++: its architecture is model-agnostic and remains robust even under relatively small-scale backend. This suggests that CoEditor++ can adapt to different computational budgets and deployment environments.

In practice, this means CoEditor++ can be scaled down for lightweight use cases without compromising structural integrity. It can also be paired with large-scale backend to enable advanced editing capabilities. The design choice to separate reasoning from generation not only enhances interpretability but also allows for flexible replacement of components as multimodal models evolve. This makes CoEditor++ a robust and adaptable solution for future model variants.

\subsection{Model Complexity}

\begin{table}[!t]
\centering
\renewcommand\arraystretch{1.2}
\caption{\textbf{Average inference time (seconds) and relative ratios for Flux-Inpainting and CoEditor++.} CoEditor++ incurs only moderate computational overhead due to its two-stage cognitive framework, remaining competitive for real-world applications.}
\sc
\resizebox{0.35\textwidth}{!}{
  \setlength{\tabcolsep}{5mm}
  \begin{tabular}{l|cc}
    \hline
    \rowcolor{lightgray}
    \textbf{Method} & \textbf{Time} & \textbf{Ratio} \\
    \hline\hline
    Flux-Inpainting & 53.85 & 100\% \\
    \hline
    \rowcolor{tinygray}CoEditor++ (Ours) & 66.35 & 123.21\% \\
    LCP & 5.18 & 9.62\% \\
    \rowcolor{tinygray}MCP & 61.17 & 113.59\% \\
    \hline
  \end{tabular}
}
\label{tab:time_ratio}
\end{table}

Table~\ref{tab:time_ratio} outlines the computational overhead introduced by CoEditor++ relative to the Flux-Inpainting baseline. Although the additional cognitive processes (LCP and MCP) introduce a modest computational increase of 23.21\%, the overall inference remains efficient and practical for real-world applications. Optimization techniques, such as pruning and quantization, can further reduce this overhead.

\subsection{Broader Impact and Limitations}

CoEditor++ significantly advances instruction-based image editing by providing precise localization, robust semantic reasoning, controlled modification, and transparent interpretability. It holds promise for diverse applications, from content moderation and creative design to ethical and compliance-oriented tasks. However, several limitations warrant further exploration. First, while reflective self-selection enhances robustness, it increases computational demands, potentially limiting scalability for resource-constrained environments. Second, although CoEditor++ performs well with complex instructions, it may struggle with extremely ambiguous or contradictory user prompts, necessitating further user interaction or clarification mechanisms. Therefore, extending this framework to dynamic scenarios such as video editing or real-time interactive settings represents a compelling direction for future research.
\section{Conclusion}

We proposed CoEditor++, which advances the frontier of instruction-based image editing by introducing a cognitively structured, fully open-source framework that explicitly separates ``\textit{what to edit}'' from ``\textit{how to edit}'' through a two-stage reasoning process. Extensive experiments and ablation studies demonstrate that the framework’s superior performance stems from its structured cognitive coordination, rather than any single model or component. CoEditor++ not only achieves state-of-the-art results across both general and responsible editing tasks, but also offers interpretability and strong generalization to complex or ambiguous scenarios. We believe this work sets a new paradigm for cognitively guided, modular visual editing, and lays the foundation for more transparent, trustworthy, and adaptable multimodal intelligence systems.

\bibliographystyle{IEEEtranN}
\bibliography{reference.bib}

\begin{thebibliography}{55}
\providecommand{\natexlab}[1]{#1}
\providecommand{\url}[1]{#1}
\csname url@samestyle\endcsname
\providecommand{\newblock}{\relax}
\providecommand{\bibinfo}[2]{#2}
\providecommand{\BIBentrySTDinterwordspacing}{\spaceskip=0pt\relax}
\providecommand{\BIBentryALTinterwordstretchfactor}{4}
\providecommand{\BIBentryALTinterwordspacing}{\spaceskip=\fontdimen2\font plus
\BIBentryALTinterwordstretchfactor\fontdimen3\font minus
  \fontdimen4\font\relax}
\providecommand{\BIBforeignlanguage}[2]{{%
\expandafter\ifx\csname l@#1\endcsname\relax
\typeout{** WARNING: IEEEtranN.bst: No hyphenation pattern has been}%
\typeout{** loaded for the language `#1'. Using the pattern for}%
\typeout{** the default language instead.}%
\else
\language=\csname l@#1\endcsname
\fi
#2}}
\providecommand{\BIBdecl}{\relax}
\BIBdecl

\bibitem[Cai et~al.(2025)Cai, Yang, Liu, Wang, Feng, Zhang, and
  Poria]{cai2025pixel}
D.~Cai, X.~Yang, Y.~Liu, D.~Wang, S.~Feng, Y.~Zhang, and S.~Poria,
  ``Pixel-level reasoning segmentation via multi-turn conversations,''
  \emph{arXiv preprint arXiv:2502.09447}, 2025.

\bibitem[Wang et~al.(2025{\natexlab{a}})Wang, Fang, Kong, Li, Xu, Yang, Li,
  Zhu, and Wang]{wang2025pixelthink}
S.~Wang, G.~Fang, L.~Kong, X.~Li, J.~Xu, S.~Yang, Q.~Li, J.~Zhu, and X.~Wang,
  ``Pixelthink: Towards efficient chain-of-pixel reasoning,'' \emph{arXiv
  preprint arXiv:2505.23727}, 2025.

\bibitem[Su et~al.(2025)Su, Wang, Ren, Lin, and Chen]{su2025pixel}
A.~Su, H.~Wang, W.~Ren, F.~Lin, and W.~Chen, ``Pixel reasoner: Incentivizing
  pixel-space reasoning with curiosity-driven reinforcement learning,''
  \emph{arXiv preprint arXiv:2505.15966}, 2025.

\bibitem[Team et~al.(2025)Team, Du, Gao, Xing, Jiang, Chen, Li, Xiao, Du, Liao,
  et~al.]{team2025kimi}
K.~Team, A.~Du, B.~Gao, B.~Xing, C.~Jiang, C.~Chen, C.~Li, C.~Xiao, C.~Du,
  C.~Liao \emph{et~al.}, ``Kimi k1. 5: Scaling reinforcement learning with
  llms, 2025,'' \emph{URL https://arxiv. org/abs/2501.12599}, 2025.

\bibitem[Chris et~al.()Chris, Peng, Wang, Qiu, Shen, Xie, Pei, Zhang, Hao,
  Song, et~al.]{chris2504skywork}
Y.~W. Chris, Y.~Peng, X.~Wang, W.~Qiu, W.~Shen, T.~Xie, J.~Pei, J.~Zhang,
  Y.~Hao, X.~Song \emph{et~al.}, ``Skywork r1v2: Multimodal hybrid
  reinforcement learning for reasoning, 2025,'' \emph{URL https://arxiv.
  org/abs/2504.16656}.

\bibitem[Jiaqi et~al.()Jiaqi, Lin, Cheng, and Shou]{wang2025think}
W.~Jiaqi, K.~Q. Lin, J.~Cheng, and M.~Z. Shou, ``Think or not? selective
  reasoning via reinforcement learning for vision-language models,'' in
  \emph{The Exploration in AI Today Workshop at ICML 2025}.

\bibitem[Wang et~al.(2025{\natexlab{b}})Wang, Zhao, Zhang, Cao, Zhan, Duan, Lu,
  Fu, Chen, Zhao, et~al.]{wang2025ovis}
G.-H. Wang, S.~Zhao, X.~Zhang, L.~Cao, P.~Zhan, L.~Duan, S.~Lu, M.~Fu, X.~Chen,
  J.~Zhao \emph{et~al.}, ``Ovis-u1 technical report,'' \emph{arXiv preprint
  arXiv:2506.23044}, 2025.

\bibitem[Chen et~al.(2025)Chen, Zhong, Li, and Yang]{chen2025unicode}
Y.~Chen, H.~Zhong, Y.~Li, and Z.~Yang, ``Cascaded large-scale
  codebooks for unified multimodal understanding and generation,'' \emph{arXiv
  preprint arXiv:2506.20214}, 2025.

\bibitem[AI et~al.(2025)AI, Gong, Zou, Zheng, Zhou, Yan, Jin, Shen, Zheng,
  Wang, et~al.]{ai2025ming}
I.~AI, B.~Gong, C.~Zou, C.~Zheng, C.~Zhou, C.~Yan, C.~Jin, C.~Shen, D.~Zheng,
  F.~Wang \emph{et~al.}, ``Ming-omni: A unified multimodal model for perception
  and generation,'' \emph{arXiv preprint arXiv:2506.09344}, 2025.

\bibitem[Fang et~al.(2025)Fang, Duan, Wang, Huang, Li, Yan, Tian, Zeng, Zhao,
  Dai, et~al.]{fang2025got}
R.~Fang, C.~Duan, K.~Wang, L.~Huang, H.~Li, S.~Yan, H.~Tian, X.~Zeng, R.~Zhao,
  J.~Dai \emph{et~al.}, ``Got: Unleashing reasoning capability of multimodal
  large language model for visual generation and editing,'' \emph{arXiv
  preprint arXiv:2503.10639}, 2025.

\bibitem[Wei et~al.(2024)Wei, Xiong, Ren, Du, Zhang, and Chen]{wei2024omniedit}
C.~Wei, Z.~Xiong, W.~Ren, X.~Du, G.~Zhang, and W.~Chen, ``Omniedit: Building
  image editing generalist models through specialist supervision,'' in
  \emph{The Thirteenth International Conference on Learning Representations},
  2024.

\bibitem[Liu et~al.(2025)Liu, Han, Xing, Yin, Wang, Cheng, Liao, Wang, Fu, Han,
  et~al.]{liu2025step1x}
S.~Liu, Y.~Han, P.~Xing, F.~Yin, R.~Wang, W.~Cheng, J.~Liao, Y.~Wang, H.~Fu,
  C.~Han \emph{et~al.}, ``Step1x-edit: A practical framework for general image
  editing,'' \emph{arXiv preprint arXiv:2504.17761}, 2025.

\bibitem[Venkatesh et~al.()Venkatesh, Dunlop, and Yanardag]{venkatesh2025crea}
K.~Venkatesh, C.~Dunlop, and P.~Yanardag, ``Crea: A collaborative multi-agent
  framework for creative image editing and generation,'' in \emph{The
  Thirty-ninth Annual Conference on Neural Information Processing Systems}.

\bibitem[Wang et~al.(2024{\natexlab{a}})Wang, Li, Yu, Wang, Li, and
  Jin]{wang2024moderator}
P.~Wang, Q.~Li, L.~Yu, Z.~Wang, A.~Li, and H.~Jin, ``Moderator: Moderating
  text-to-image diffusion models through fine-grained context-based policies,''
  in \emph{Proceedings of the 2024 on ACM SIGSAC Conference on Computer and
  Communications Security}, 2024, pp. 1181--1195.

\bibitem[Chang et~al.(2024)Chang, Liu, Zhang, and Guo]{chang2024editscribe}
R.-C. Chang, Y.~Liu, L.~Zhang, and A.~Guo, ``Editscribe: Non-visual image
  editing with natural language verification loops,'' in \emph{Proceedings of
  the 26th International ACM SIGACCESS Conference on Computers and
  Accessibility}, 2024, pp. 1--19.

\bibitem[Meding and Sorge(2025)]{meding2025constitutes}
K.~Meding and C.~Sorge, ``What constitutes a deep fake? the blurry line between
  legitimate processing and manipulation under the eu ai act,'' in
  \emph{Proceedings of the 2025 Symposium on Computer Science and Law}, 2025,
  pp. 152--159.

\bibitem[Ni et~al.(2024{\natexlab{a}})Ni, Shen, Zhang, and
  Zuo]{ni2024responsible}
M.~Ni, Y.~Shen, L.~Zhang, and W.~Zuo, ``Responsible visual editing,'' in
  \emph{European Conference on Computer Vision}.\hskip 1em plus 0.5em minus
  0.4em\relax Springer, 2024, pp. 314--330.

\bibitem[Wang et~al.(2025{\natexlab{c}})Wang, Zhang, Bai, Zhao, Liu, and
  Tu]{wang2025edit}
H.~Wang, Y.~Zhang, R.~Bai, Y.~Zhao, S.~Liu, and Z.~Tu, ``Edit away and my face
  will not stay: Personal biometric defense against malicious generative
  editing,'' in \emph{Proceedings of the Computer Vision and Pattern
  Recognition Conference}, 2025, pp. 23\,806--23\,816.

\bibitem[Xu et~al.(2025)Xu, Kong, Wang, Pan, Lin, and Liu]{xu2025insightedit}
Y.~Xu, J.~Kong, J.~Wang, X.~Pan, B.~Lin, and Q.~Liu, ``Insightedit: Towards
  better instruction following for image editing,'' in \emph{Proceedings of the
  Computer Vision and Pattern Recognition Conference}, 2025, pp. 2694--2703.

\bibitem[Zhang et~al.(2025)Zhang, Xie, Lu, Yang, and Yang]{zhang2025context}
Z.~Zhang, J.~Xie, Y.~Lu, Z.~Yang, and Y.~Yang, ``In-context edit: Enabling
  instructional image editing with in-context generation in large scale
  diffusion transformer,'' \emph{arXiv preprint arXiv:2504.20690}, 2025.

\bibitem[Ma et~al.(2025)Ma, Peng, Guo, Chen, Lu, and Yang]{ma2025x2i}
J.~Ma, Q.~Peng, X.~Guo, C.~Chen, H.~Lu, and Z.~Yang, ``X2i: Seamless
  integration of multimodal understanding into diffusion transformer via
  attention distillation,'' \emph{arXiv preprint arXiv:2503.06134}, 2025.

\bibitem[Huang et~al.(2024)Huang, Xie, Wang, Yuan, Cun, Ge, Zhou, Dong, Huang,
  Zhang, et~al.]{huang2024smartedit}
Y.~Huang, L.~Xie, X.~Wang, Z.~Yuan, X.~Cun, Y.~Ge, J.~Zhou, C.~Dong, R.~Huang,
  R.~Zhang \emph{et~al.}, ``Smartedit: Exploring complex instruction-based
  image editing with multimodal large language models,'' in \emph{Proceedings
  of the IEEE/CVF Conference on Computer Vision and Pattern Recognition}, 2024,
  pp. 8362--8371.

\bibitem[Hurst et~al.(2024)Hurst, Lerer, Goucher, Perelman, Ramesh, Clark,
  Ostrow, Welihinda, Hayes, Radford, et~al.]{hurst2024gpt}
A.~Hurst, A.~Lerer, A.~P. Goucher, A.~Perelman, A.~Ramesh, A.~Clark, A.~Ostrow,
  A.~Welihinda, A.~Hayes, A.~Radford \emph{et~al.}, ``Gpt-4o system card,''
  \emph{arXiv preprint arXiv:2410.21276}, 2024.

\bibitem[{Google}(2025)]{Nano_Banana}
{Google}, ``Nano banana,'' \url{https://gemini.google.com}, 2025, accessed:
  2025‑11‑14.

\bibitem[Google(2025)]{Nano_Banana_Pro}
Google, ``Nano banana pro,'' \url{https://gemini.google.com}, 2025, accessed:
  2025‑12‑11.

\bibitem[Ge et~al.(2024)Ge, Zhao, Zhu, Ge, Yi, Song, Li, Ding, and
  Shan]{ge2024seed}
Y.~Ge, S.~Zhao, J.~Zhu, Y.~Ge, K.~Yi, L.~Song, C.~Li, X.~Ding, and Y.~Shan,
  ``Seed-x: Multimodal models with unified multi-granularity comprehension and
  generation,'' \emph{arXiv preprint arXiv:2404.14396}, 2024.

\bibitem[Wu et~al.(2025)Wu, Li, Zhou, Lin, Gao, Yan, Yin, Bai, Xu, Chen,
  et~al.]{wu2025qwen}
C.~Wu, J.~Li, J.~Zhou, J.~Lin, K.~Gao, K.~Yan, S.-m. Yin, S.~Bai, X.~Xu,
  Y.~Chen \emph{et~al.}, ``Qwen-image technical report,'' \emph{arXiv preprint
  arXiv:2508.02324}, 2025.

\bibitem[Kingma and Welling(2013)]{kingma2013auto}
D.~P. Kingma and M.~Welling, ``Auto-encoding variational bayes,'' \emph{arXiv
  preprint arXiv:1312.6114}, 2013.

\bibitem[Goodfellow et~al.(2014)Goodfellow, Pouget-Abadie, Mirza, Xu,
  Warde-Farley, Ozair, Courville, and Bengio]{goodfellow2014generative}
I.~J. Goodfellow, J.~Pouget-Abadie, M.~Mirza, B.~Xu, D.~Warde-Farley, S.~Ozair,
  A.~Courville, and Y.~Bengio, ``Generative adversarial nets,'' \emph{Advances
  in neural information processing systems}, vol.~27, 2014.

\bibitem[Croitoru et~al.(2023)Croitoru, Hondru, Ionescu, and
  Shah]{croitoru2023diffusion}
F.-A. Croitoru, V.~Hondru, R.~T. Ionescu, and M.~Shah, ``Diffusion models in
  vision: A survey,'' \emph{IEEE transactions on pattern analysis and machine
  intelligence}, vol.~45, no.~9, pp. 10\,850--10\,869, 2023.

\bibitem[Brooks et~al.(2023)Brooks, Holynski, and Efros]{Brooks_2023_CVPR}
T.~Brooks, A.~Holynski, and A.~A. Efros, ``Instructpix2pix: Learning to follow
  image editing instructions,'' in \emph{Proceedings of the IEEE/CVF Conference
  on Computer Vision and Pattern Recognition (CVPR)}, June 2023, pp.
  18\,392--18\,402.

\bibitem[Ho and Salimans(2022)]{ho2022classifier}
J.~Ho and T.~Salimans, ``Classifier-free diffusion guidance,'' \emph{arXiv
  preprint arXiv:2207.12598}, 2022.

\bibitem[Zhang et~al.(2023)Zhang, Mo, Chen, Sun, and Su]{zhang2023magicbrush}
K.~Zhang, L.~Mo, W.~Chen, H.~Sun, and Y.~Su, ``Magicbrush: A manually annotated
  dataset for instruction-guided image editing,'' \emph{Advances in Neural
  Information Processing Systems}, vol.~36, pp. 31\,428--31\,449, 2023.

\bibitem[Hui et~al.(2025)Hui, Yang, Zhao, Shi, Wang, Wang, Xie, and
  Zhou]{hui2025hq}
M.~Hui, S.~Yang, B.~Zhao, Y.~Shi, H.~Wang, P.~Wang, C.~Xie, and Y.~Zhou,
  ``Hq-edit: A high-quality dataset for instruction-based image editing,'' in
  \emph{ICLR}, 2025.

\bibitem[Zhao et~al.(2024)Zhao, Ma, Chen, Si, Wu, An, Yu, Zhang, Li, and
  Chang]{zhao2024ultraedit}
H.~Zhao, X.~S. Ma, L.~Chen, S.~Si, R.~Wu, K.~An, P.~Yu, M.~Zhang, Q.~Li, and
  B.~Chang, ``Ultraedit: Instruction-based fine-grained image editing at
  scale,'' \emph{Advances in Neural Information Processing Systems}, vol.~37,
  pp. 3058--3093, 2024.

\bibitem[Yu et~al.(2025{\natexlab{a}})Yu, Chow, Yue, Pan, Wu, Wan, Li, Tang,
  Zhang, and Zhuang]{yu2025anyedit}
Q.~Yu, W.~Chow, Z.~Yue, K.~Pan, Y.~Wu, X.~Wan, J.~Li, S.~Tang, H.~Zhang, and
  Y.~Zhuang, ``Anyedit: Mastering unified high-quality image editing for any
  idea,'' in \emph{Proceedings of the Computer Vision and Pattern Recognition
  Conference}, 2025, pp. 26\,125--26\,135.

\bibitem[Fu et~al.(2023)Fu, Hu, Du, Wang, Yang, and Gan]{fu2023guiding}
T.-J. Fu, W.~Hu, X.~Du, W.~Y. Wang, Y.~Yang, and Z.~Gan, ``Guiding
  instruction-based image editing via multimodal large language models,''
  \emph{arXiv preprint arXiv:2309.17102}, 2023.

\bibitem[Xiao et~al.(2025)Xiao, Wang, Zhou, Yuan, Xing, Yan, Li, Wang, Huang,
  and Liu]{xiao2025omnigen}
S.~Xiao, Y.~Wang, J.~Zhou, H.~Yuan, X.~Xing, R.~Yan, C.~Li, S.~Wang, T.~Huang,
  and Z.~Liu, ``Omnigen: Unified image generation,'' in \emph{Proceedings of
  the Computer Vision and Pattern Recognition Conference}, 2025, pp.
  13\,294--13\,304.

\bibitem[Liu et~al.(2023)Liu, Li, Wu, and Lee]{liu2023visual}
H.~Liu, C.~Li, Q.~Wu, and Y.~J. Lee, ``Visual instruction tuning,''
  \emph{Advances in neural information processing systems}, vol.~36, pp.
  34\,892--34\,916, 2023.

\bibitem[Bai et~al.(2025)Bai, Chen, Liu, Wang, Ge, Song, Dang, Wang, Wang,
  Tang, et~al.]{bai2025qwen2}
S.~Bai, K.~Chen, X.~Liu, J.~Wang, W.~Ge, S.~Song, K.~Dang, P.~Wang, S.~Wang,
  J.~Tang \emph{et~al.}, ``Qwen2. 5-vl technical report,'' \emph{arXiv preprint
  arXiv:2502.13923}, 2025.

\bibitem[Wang et~al.(2024{\natexlab{b}})Wang, Bai, Tan, Wang, Fan, Bai, Chen,
  Liu, Wang, Ge, et~al.]{wang2024qwen2}
P.~Wang, S.~Bai, S.~Tan, S.~Wang, Z.~Fan, J.~Bai, K.~Chen, X.~Liu, J.~Wang,
  W.~Ge \emph{et~al.}, ``Qwen2-vl: Enhancing vision-language model's perception
  of the world at any resolution,'' \emph{arXiv preprint arXiv:2409.12191},
  2024.

\bibitem[Chen et~al.(2024)Chen, Wang, Cao, Liu, Gao, Cui, Zhu, Ye, Tian, Liu,
  et~al.]{chen2024expanding}
Z.~Chen, W.~Wang, Y.~Cao, Y.~Liu, Z.~Gao, E.~Cui, J.~Zhu, S.~Ye, H.~Tian,
  Z.~Liu \emph{et~al.}, ``Expanding performance boundaries of open-source
  multimodal models with model, data, and test-time scaling,'' \emph{arXiv
  preprint arXiv:2412.05271}, 2024.

\bibitem[Zhu et~al.(2025)Zhu, Wang, Chen, Liu, Ye, Gu, Tian, Duan, Su, Shao,
  et~al.]{zhu2025internvl3}
J.~Zhu, W.~Wang, Z.~Chen, Z.~Liu, S.~Ye, L.~Gu, H.~Tian, Y.~Duan, W.~Su,
  J.~Shao \emph{et~al.}, ``Internvl3: Exploring advanced training and test-time
  recipes for open-source multimodal models,'' \emph{arXiv preprint
  arXiv:2504.10479}, 2025.

\bibitem[Wei et~al.(2022)Wei, Wang, Schuurmans, Bosma, Xia, Chi, Le, Zhou,
  et~al.]{wei2022chain}
J.~Wei, X.~Wang, D.~Schuurmans, M.~Bosma, F.~Xia, E.~Chi, Q.~V. Le, D.~Zhou
  \emph{et~al.}, ``Chain-of-thought prompting elicits reasoning in large
  language models,'' \emph{Advances in neural information processing systems},
  vol.~35, pp. 24\,824--24\,837, 2022.

\bibitem[Guo et~al.(2025)Guo, Yang, Zhang, Song, Zhang, Xu, Zhu, Ma, Wang, Bi,
  et~al.]{guo2025deepseek}
D.~Guo, D.~Yang, H.~Zhang, J.~Song, R.~Zhang, R.~Xu, Q.~Zhu, S.~Ma, P.~Wang,
  X.~Bi \emph{et~al.}, ``Deepseek-r1: Incentivizing reasoning capability in
  llms via reinforcement learning,'' \emph{arXiv preprint arXiv:2501.12948},
  2025.

\bibitem[Yu et~al.(2025{\natexlab{b}})Yu, Zhang, Zhu, Yuan, Zuo, Yue, Dai, Fan,
  Liu, Liu, et~al.]{yu2025dapo}
Q.~Yu, Z.~Zhang, R.~Zhu, Y.~Yuan, X.~Zuo, Y.~Yue, W.~Dai, T.~Fan, G.~Liu,
  L.~Liu \emph{et~al.}, ``Dapo: An open-source llm reinforcement learning
  system at scale,'' \emph{arXiv preprint arXiv:2503.14476}, 2025.

\bibitem[Yue et~al.(2025)Yue, Yuan, Yu, Zuo, Zhu, Xu, Chen, Wang, Fan, Du,
  et~al.]{yue2025vapo}
Y.~Yue, Y.~Yuan, Q.~Yu, X.~Zuo, R.~Zhu, W.~Xu, J.~Chen, C.~Wang, T.~Fan, Z.~Du
  \emph{et~al.}, ``Vapo: Efficient and reliable reinforcement learning for
  advanced reasoning tasks,'' \emph{arXiv preprint arXiv:2504.05118}, 2025.

\bibitem[Shen et~al.(2025)Shen, Liu, Li, Fang, Ma, Liao, Shen, Zhang, Zhao,
  Zhang, et~al.]{shen2025vlm}
H.~Shen, P.~Liu, J.~Li, C.~Fang, Y.~Ma, J.~Liao, Q.~Shen, Z.~Zhang, K.~Zhao,
  Q.~Zhang \emph{et~al.}, ``Vlm-r1: A stable and generalizable r1-style large
  vision-language model,'' \emph{arXiv preprint arXiv:2504.07615}, 2025.

\bibitem[Huang et~al.(2025)Huang, Jia, Zhai, Cao, Ye, Zhao, Xu, Hu, and
  Lin]{huang2025vision}
W.~Huang, B.~Jia, Z.~Zhai, S.~Cao, Z.~Ye, F.~Zhao, Z.~Xu, Y.~Hu, and S.~Lin,
  ``Vision-r1: Incentivizing reasoning capability in multimodal large language
  models,'' \emph{arXiv preprint arXiv:2503.06749}, 2025.

\bibitem[Kang et~al.(2025)Kang, Song, Zhou, Qin, Yang, Liu, Torr, Bai, and
  Yin]{kang2025viki}
L.~Kang, X.~Song, H.~Zhou, Y.~Qin, J.~Yang, X.~Liu, P.~Torr, L.~Bai, and
  Z.~Yin, ``Viki-r: Coordinating embodied multi-agent cooperation via
  reinforcement learning,'' \emph{arXiv preprint arXiv:2506.09049}, 2025.

\bibitem[Ni et~al.(2024{\natexlab{b}})Ni, Wu, Wang, Yin, Wang, Liu, and
  Duan]{ni2024ores}
M.~Ni, C.~Wu, X.~Wang, S.~Yin, L.~Wang, Z.~Liu, and N.~Duan, ``Ores:
  Open-vocabulary responsible visual synthesis,'' in \emph{Proceedings of the
  AAAI Conference on Artificial Intelligence}, vol.~38, no.~19, 2024, pp.
  21\,473--21\,481.

\bibitem[Hore and Ziou(2010)]{hore2010image}
A.~Hore and D.~Ziou, ``Image quality metrics: Psnr vs. ssim,'' in \emph{2010
  20th international conference on pattern recognition}.\hskip 1em plus 0.5em
  minus 0.4em\relax IEEE, 2010, pp. 2366--2369.

\bibitem[Zhang et~al.(2018)Zhang, Isola, Efros, Shechtman, and
  Wang]{zhang2018unreasonable}
R.~Zhang, P.~Isola, A.~A. Efros, E.~Shechtman, and O.~Wang, ``The unreasonable
  effectiveness of deep features as a perceptual metric,'' in \emph{Proceedings
  of the IEEE conference on computer vision and pattern recognition}, 2018, pp.
  586--595.

\bibitem[Radford et~al.(2021)Radford, Kim, Hallacy, Ramesh, Goh, Agarwal,
  Sastry, Askell, Mishkin, Clark, et~al.]{radford2021learning}
A.~Radford, J.~W. Kim, C.~Hallacy, A.~Ramesh, G.~Goh, S.~Agarwal, G.~Sastry,
  A.~Askell, P.~Mishkin, J.~Clark \emph{et~al.}, ``Learning transferable visual
  models from natural language supervision,'' in \emph{International conference
  on machine learning}.\hskip 1em plus 0.5em minus 0.4em\relax PmLR, 2021, pp.
  8748--8763.

\bibitem[Geng et~al.(2024)Geng, Yang, Hang, Li, Gu, Zhang, Bao, Zhang, Li, Hu,
  et~al.]{geng2024instructdiffusion}
Z.~Geng, B.~Yang, T.~Hang, C.~Li, S.~Gu, T.~Zhang, J.~Bao, Z.~Zhang, H.~Li,
  H.~Hu \emph{et~al.}, ``Instructdiffusion: A generalist modeling interface for
  vision tasks,'' in \emph{Proceedings of the IEEE/CVF Conference on computer
  vision and pattern recognition}, 2024, pp. 12\,709--12\,720.

\end{thebibliography}
%

\end{document}